\documentclass[fleqn,usenatbib]{mnras}

\usepackage{newtxtext,newtxmath}

\usepackage[T1]{fontenc}

\DeclareRobustCommand{\VAN}[3]{#2}
\let\VANthebibliography\thebibliography
\def\thebibliography{\DeclareRobustCommand{\VAN}[3]{##3}\VANthebibliography}

\usepackage{graphicx}	
\graphicspath{ {images/} }
\usepackage{amsmath}	

\usepackage[dvipsnames]{xcolor}

\usepackage{xspace}

\newcommand{\Msun}{$\rm\,M_\odot$\xspace}
\newcommand{\Lsun}{$\rm\,L_\odot$\xspace}
\newcommand{\kms}{$\rm\,km\,s^{-1}$\xspace}
\newcommand{\pc}{$\rm\,pc$\xspace}
\newcommand{\kpc}{$\rm\,kpc$\xspace}
\newcommand{\Gyr}{$\rm\,Gyr$\xspace}
\newcommand{\degree}{$^\circ$\xspace}
\newcommand{\Arcmin}{$\rm\,arcmin$\xspace}

\newcommand{\dex}{$\rm\,dex$\xspace}
\newcommand{\LCDM}{$\rm\Lambda CDM$\xspace}
\newcommand{\about}{$\sim$}

\defcitealias{battaglia_study_2011}{B11}
\defcitealias{di_cintio_mass-dependent_2014}{DC14}
\defcitealias{dutton_cold_2014}{DM14}
\defcitealias{mcmillan_mass_2017}{M17}

\title[Recovering the infall mass for Sextans]{Recovering the infall mass for Milky Way satellite galaxy Sextans}

\author[Tian et al.]{
Tingting Tian,$^{1}$
Jiang Chang,$^{3}$
Go Ogiya,$^{1}$
Xi Kang,$^{1,2,3}$\thanks{Corresponding author, E-mail: kangxi@zju.edu.cn}
Renyue Cen,$^{2,1}$
\\
$^{1}$Institute of Astronomy, School of Physics, Zhejiang University, Hangzhou 310027, China\\
$^{2}$Center for Cosmology and Computational Astrophysics, Zhejiang University, Hangzhou 310027, China\\
$^{3}$Purple Mountain Observatory, 10 Yuan hua Road, Nanjing 210034, China
}

\date{Accepted XXX. Received YYY; in original form ZZZ}

\pubyear{2026}

\begin{document}
\label{firstpage}
\pagerange{\pageref{firstpage}--\pageref{lastpage}}
\maketitle

\begin{abstract}
Understanding the formation and evolution of the Milky Way (MW) requires detailed knowledge of its satellite galaxies. In this study, we focus on the Sextans dwarf spheroidal (dSph) galaxy, a faint, dark matter (DM)-dominated satellite, to investigate the role of tidal and baryonic effects in shaping its observed properties. Using tailored $N$-body simulations, we explore possible orbits of Sextans in different MW models to reconstruct its progenitor's properties. Our simulations demonstrate the stars in Sextans are only mildly affected by galactic tides and the stellar kinematics provide robust constraints on its dynamical mass within the half-light radius, while the tidal mass loss of its DM component depends primarily on MW mass. The recovered infall mass of Sextans ranges from $1.22$ to $3.14\times10^9$\Msun for MW masses from $0.8$ to $2\times10^{12}$\Msun. If the DM density remained as cuspy as NFW profile, the infall mass would be smaller by a factor of 2. Although with large ranges, the possible infall masses of Sextans recovered by our simulations are consistent with the stellar mass-halo mass relation in TNG50 and abundance matching results. We find some cases for the cuspy DM density profile where the infall mass is smaller than $10^9$\Msun, possibly indicating that star formation in Sextans is more efficient than in other satellites. The recovered DM halo structural parameters from our simulations provide valuable constraints for future studies on the DM content and formation history of Sextans.
\end{abstract}



\section{Introduction}
\label{sec:Introduction}
The MW serves as a unique local laboratory for testing predictions of the cold dark matter (CDM) model on small scales, with many of the strongest constraints coming from its dwarf satellite galaxies \citep[e.g.,][]{bullock_small-scale_2017,simon_faintest_2019}. To date, over sixty satellite galaxies have been identified through deep and wide surveys \citep[e.g.,][and references therein]{simon_faintest_2019,doliva-dolinsky_satellite_2025}. However, due to their faintness, reliable measurements of key properties such as luminosity, half-light radius, and line-of-sight (LOS) velocity dispersion are only available for the bright and closer satellites (e.g., \citealt{wolf_accurate_2010,mcconnachie_observed_2012}, see \citealt{mcconnachie_revised_2020} for more references).

The advent of Gaia data releases, particularly \textit{Gaia} data release 2 \citep[DR2,][]{gaia_collaboration_gaia_2018} and \textit{Gaia} Early Data Release 3 \citep[EDR3,][]{gaia_collaboration_gaia_2021}, has enabled precise proper motion (PM) measurements for most satellites in and around the MW \citep[e.g.][]{fritz_gaia_2018,mcconnachie_revised_2020,li_gaia_2021,battaglia_gaia_2022,pace_proper_2022}. With accurate 3D positions and velocities, it is now possible to reconstruct the orbital evolution of these satellites using models of the MW's mass distribution \citep{gaia_collaboration_gaia_2018,fritz_gaia_2018,simon_gaia_2018,patel_orbital_2020,li_gaia_2021,battaglia_gaia_2022,pace_proper_2022}. A handful of satellites were found to be near at their pericenter \citep[e.g.][]{fritz_gaia_2018,simon_gaia_2018,li_gaia_2021,pace_proper_2022}, indicating possible tidal stripping or disruption among these systems. However, these studies often treat satellites as point masses, neglecting their internal structures.

To constrain the internal structures of satellites and the tidal effects they have experienced, tailored $N$-body simulations are often employed to reproduce the current properties of satellites, such as stellar mass, half-light radius, and velocity dispersion, by modeling their evolution from infall to the present day based on observationally motivated orbits. This approach has been successfully applied to several satellite galaxies, including Leo I \citep{sohn_exploring_2007}, Carina \citep{munoz_modeling_2008,ural_low_2015}, Fornax \citep{battaglia_effect_2015} and Crater II \citep{sanders_tidal_2018}, particularly in the pre-\textit{Gaia} DR2 era. Post-\textit{Gaia} DR2, similar techniques have been applied to Sculptor \citep{iorio_effect_2019}, Canes Venatici I \citep{matus_carrillo_modelling_2020}, Fornax \citep{borukhovetskaya_tidal_2021,di_cintio_tidal_2024} and Crater II \citep{borukhovetskaya_galactic_2022,zhang_self-interacting_2024}, leveraging more accurate PMs to reconstruct their orbits.

Understanding the DM halo mass and the structural properties of satellite galaxies would provide critical insights into several astrophysical questions. For example, the stellar mass-halo mass (SMHM) relation at the low-mass end is still uncertain and depends on the galaxy formation model \citep{sales_baryonic_2022}, and accurate measurements of satellite halo masses can help constrain this relation and its scatter. These properties of MW satellite galaxies, along with field dwarfs, can be used further to constrain how the Local Group (LG) environment affects star formation efficiency in dwarf galaxies \citep[e.g.,][]{santistevan_formation_2020,grand_overview_2024}. Besides, several small-scale challenges to the \LCDM model are closely related to the DM halo mass and inner density of MW satellite galaxies \citep[e.g.][]{bullock_small-scale_2017}. For example, the too-big-to-fail (TBTF) problem \citep{boylan-kolchin_too_2011,boylan-kolchin_milky_2012}, which at first suggests a deficit of massive DM subhalos in the MW, can be alleviated by lowering the MW mass, the scatter of subhalo mass function, baryonic feedback, and tidal effects \citep[see][for a comprehensive review]{bullock_small-scale_2017,sales_baryonic_2022}. To fully resolve the TBTF problem, a better understanding of our MW mass and the subhalo mass function is needed.

Previous simulation work has shown that for Fornax \citep{battaglia_effect_2015,borukhovetskaya_tidal_2021} and Sculptor \citep{iorio_effect_2019}, which are more luminous satellites in the MW with luminosity of $1.7\times10^7$\Lsun and $2.5\times10^6$\Lsun respectively, the progenitor DM halo mass with \about$10^9-10^{10}$\Msun can well reproduce the observed properties, which is also expected by the SMHM relation down to this scale \citep[e.g.][]{sales_baryonic_2022}. However, for the MW satellite galaxy Carina, which has a luminosity of $4.3\times10^5$\Lsun, the progenitor DM halo mass is constrained to be $3.6_{-2.3}^{+3.8}\times10^8$\Msun, such a low mass would indicate the stochasticity of galaxy formation at halo masses below $10^{10}$\Msun \citep{ural_low_2015}.

In this study, we focus on another faint MW satellite, Sextans dSph galaxy \citep{irwin_new_1990}, to recover its DM halo properties prior to infall using tailored $N$-body simulations. Sextans is spatially extended, diffuse, and DM-dominated within its half-light radius, with a total luminosity of a few $10^5$\Lsun (\citealp[hereafter B11]{walker_universal_2009,wolf_accurate_2010,battaglia_study_2011}). Jeans modeling suggests that Sextans may have a cored DM halo or a cuspy halo with low concentration \citepalias{battaglia_study_2011}. Some studies attribute the kinematically cold substructures found in the center of Sextans (\citealp{kleyna_photometrically_2004,walker_kinematic_2006};\citetalias{battaglia_study_2011}) to the remnants of star clusters, possibly indicating a cored DM halo profile \citep{lora_sextans_2013,kim_possible_2019}.

Collisionless cosmological simulations predict the DM halo density is cuspy towards the halo center \citep{navarro_structure_1996,navarro_universal_1997}. According to some studies, however, stellar feedback processes can alter the inner density profile of DM haloes. One scenario is that the potential fluctuations caused by supernovae-driven outflows transfer energy into collisionless particles, flattening the central DM density cusp into a cored profile \citep{navarro_cores_1996,gnedin_maximum_2002,pontzen_how_2012,ogiya_core-cusp_2014}, and the extent of this modification has been found to depend on the stellar-to-halo mass ratio \citep[e.g.,][]{di_cintio_dependence_2014,tollet_nihao_2016,fitts_fire_2017,freundlich_dekel-zhao_2020,lazar_dark_2020}. Besides, the star formation history (SFH) also influences the cusp-core transformation. For example, large core formation requires long period of star formation after the rapid DM halo build-up \citep{onorbe_forged_2015,chan_impact_2015} or after reionization \citep{muni_span_2024}. Apart from these, the density threshold for star formation also plays a critical role for core formation as indicated by recent studies \citep{bose_no_2019,benitez-llambay_baryon-induced_2019,jahn_real_2023}. According to these studies, there is no core formation in simulations which adopt a low density threshold for star formation.

Since the DM halo density profile of Sextans is still uncertain, we adopt the baryonic effect model presented in \citet[DC14]{di_cintio_mass-dependent_2014} to determine its DM distribution prior to infall. However, we will also explore a cuspy profile for Sextans to investigate its impacts on derived halo properties. To further understand the formation of Sextans, we compare the recovered halo masses with the SMHM relation in the TNG50-1 cosmological simulation \citep[hereafter TNG50]{nelson_illustristng_2021,nelson_first_2019,pillepich_first_2019} and a few abundance matching results to investigate the consistency of Sextans with present galaxy formation models.

The paper is organized as follows: Section~\ref{sec:Method for tailored N-body simulation} summarizes the observed properties of Sextans and details the simulation setup for our tailored method. Section~\ref{sec:Results} presents the results of our simulations. Finally, Section~\ref{sec:Discussion} discusses the uncertainties and implications of our findings, and Section~\ref{sec:Summary} provides a summary.

\section[Method for tailored N-body simulation]{Method for tailored $N$-body simulation}
\label{sec:Method for tailored N-body simulation}
This section outlines the setup of our $N$-body simulations, specifically tailored for the MW satellite galaxy Sextans dSph (hereafter Sextans). Observation constraints for Sextans are summarized in Section~\ref{sec:The observed properties of Sextans}. The model of Sextans is described in Section~\ref{sec:The satellite galaxy model}. We place the satellite on orbits determined based on its observed position and velocity and evolve it in analytic MW potentials. In this study, we adopt three distinct halo masses for the MW, allowing them to evolve with redshift. Further details are provided in Section~\ref{sec:The Milky Way model} and Section~\ref{sec:Orbital integration for Sextans}. We iteratively adjust the DM mass, profile, stellar mass, and scale radius to best reproduce the observed LOS velocity dispersion, projected half-light radius, and stellar mass of Sextans, as is described in Section~\ref{sec:Simulation set-up}.

Throughout this paper, the halo virial radius is defined as the radius within which the mean density is 200 times the critical density of the universe at a given redshift. The quantities dependent on the cosmological parameters (such as virial mass, concentration, lookback time) are calculated with $h=0.6774$, $\Omega_m=0.3089$ and $\Omega_\Lambda=0.6911$ \citep{planck_collaboration_planck_2016}. The masses of the MW components are denoted by the capital letter $M$, while those of the satellite components are denoted by the lowercase letter $m$. The capital letter $R$ denotes the 2D projected radius in the sky, while the lowercase letter $r$ denotes the physical radius from the center of the MW or satellite.

\subsection{The observed properties of Sextans}
\label{sec:The observed properties of Sextans}
\begin{figure}
    \includegraphics[width=\columnwidth]{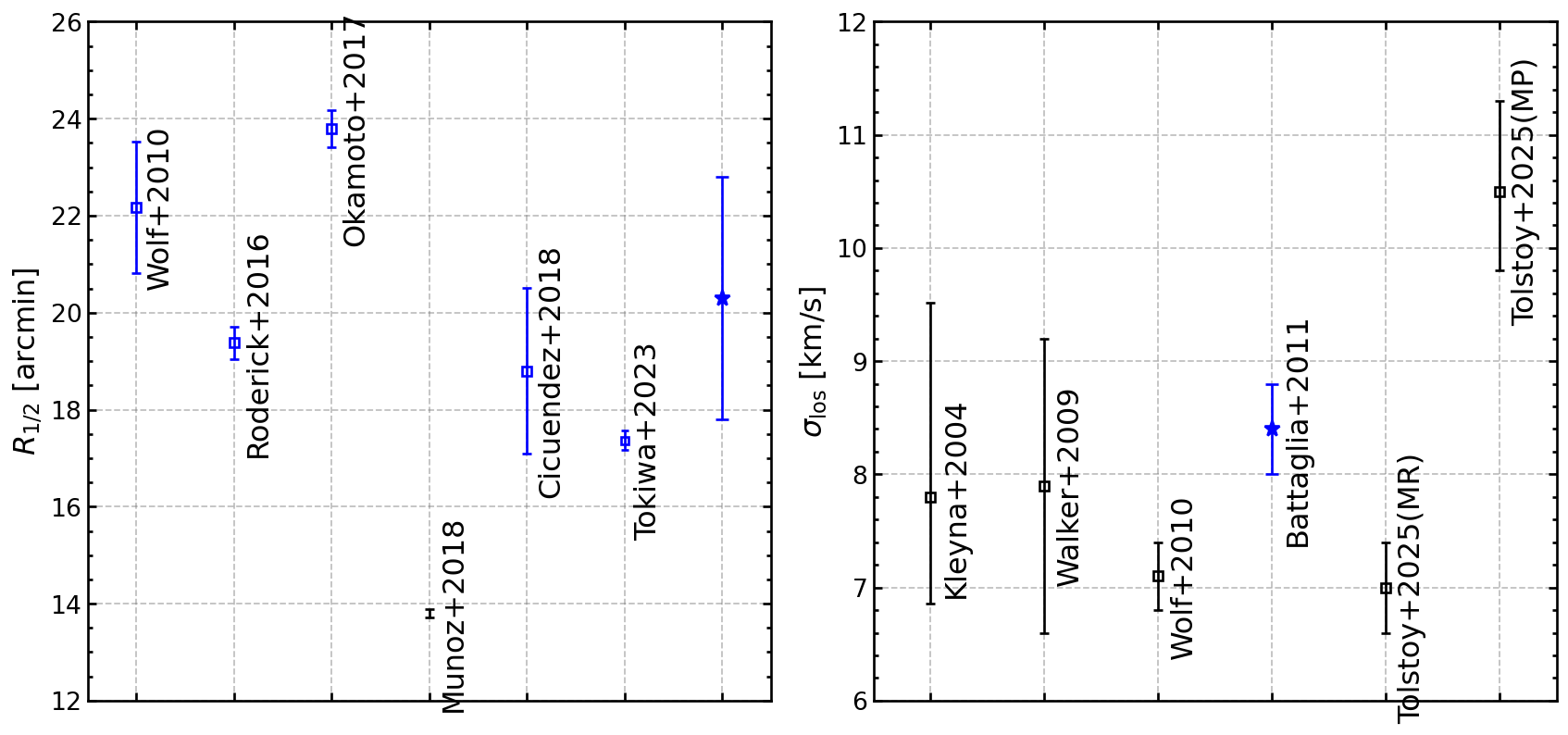}
    \caption{
    The observed half-light radius and LOS velocity dispersion of Sextans from literatures. In the left panel, we convert the elliptical radius to circular one based on the provided ellipticity in each work and the blue squares are used to calculate the average $R_{1/2}$ and its uncertainty, which is marked by the blue star. For LOS velocity dispersion, we adopt the value in \citetalias{battaglia_study_2011}, which is shown by the blue star in the right panel.
    }
    \label{fig:obsHLR-SigmaLOS}
\end{figure}
\begin{table*}
    \centering
    \caption{
    The observed properties of Sextans.
    The LOS velocity dispersion $\sigma_{\rm los}$ \citepalias{battaglia_study_2011}, stellar mass $m_{*}$ and projected half-light radius $R_{1/2}$ of Sextans are in the upper rows. Here we adopt the V-band luminosity provided in \citep{wolf_accurate_2010} and convert it to stellar mass assuming a stellar mass-to-light ratio of 1. The $R_{1/2}$ in units of pc and its 3D de-projected radius $r_{1/2}=4/3R_{1/2}$ are converted assuming a distance of 92.5\kpc. The circular velocity at $r_{1/2}$ is derived by $v_{1/2}=\sqrt{3}\sigma_{\rm los}$ \citep{wolf_accurate_2010}.
    The right ascension (R.A.), declination (Dec.), heliocentric distance \citep[$d$,][]{okamoto_population_2017} and LOS velocity \citep[$v_{\rm los}$][]{fritz_gaia_2018}, proper motions ($\mu_{\alpha,*}, \mu_{\delta}$) and their correlation coefficient $\rho_{\mu_{\alpha,*}}^{\mu_\delta}$ \citep{li_gaia_2021} of Sextans are shown in lower rows.
    }
    \label{tab:SextansObs}
    \begin{tabular}{c cc cc cc c}
    \hline
    Satellite & $\sigma_{\rm los}$ & $m_{*}$ & $R_{1/2}$ & $r_{1/2}$ & $v_{1/2}$ & & \\
    & ($\rm km\,s^{-1}$) & ($\rm 10^5\,M_{\odot}$) & (arcmin) & (pc) & ($\rm km\,s^{-1}$) & & \\
    \hline
    Sextans & $8.4\pm0.4$ & $5.9_{-1.4}^{+2.0}$ & $20.3\pm2.5$ & $728\pm91$ & $14.5\pm0.7$ & & \\
    & &  & ($546\pm67$\pc) & & & & \\
    \hline
    & R.A. & Dec. & $d$ & $v_{\rm los}$ & $\mu_{\alpha,*}$ & $\mu_\delta$ & $\rho_{\mu_{\alpha,*}}^{\mu_\delta}$ \\
    & (deg) & (deg) & (kpc) & ($\rm km\,s^{-1}$) & ($\rm mas\,yr^{-1}$) & ($\rm mas\,yr^{-1}$) & \\
    \hline
    & 153.268 & -1.620 & $92.5\pm2.2$ & $224.2\pm0.1$ & $-0.403\pm0.021$ & $0.029\pm0.021$ & -0.090 \\
    \hline
    \end{tabular}
\end{table*}
Sextans dSph is discovered by \citet{irwin_new_1990} and is one of the faintest and most diffuse classical MW satellite galaxies. Its low central brightness ($18.2\pm0.5\,$mag$\,$arcmin$^{-2}$), large extent on the sky (nominal King tidal radius $\sim$120\Arcmin) and low galactic latitude ($l=243.5$\degree, $b=+42.3$\degree) make it a difficult object to study.

Structural parameters of Sextans have been measured by a handful of work \citep{irwin_new_1990,irwin_structural_1995,roderick_structural_2016,okamoto_population_2017,munoz_megacam_2018,cicuendez_tracing_2018,tokiwa_study_2023,tolstoy_3d_2025}. Among these studies, it has been established that both the Plummer (\citealt{plummer_problem_1911}, e.g., \citealt{roderick_structural_2016}) and King (\citealt{king_structure_1962}, e.g., \citealt{roderick_structural_2016,okamoto_population_2017,cicuendez_tracing_2018}) models provide a good fit to the surface number density of Sextans. The left panel of Fig.~\ref{fig:obsHLR-SigmaLOS} shows the 2D half-light radius $R_{1/2}$ derived from these papers\footnote{Here we convert the elliptical radius to a circular one by $R_{1/2}=a_{1/2}\sqrt{1-\epsilon}$, where $a_{1/2}$ is the half-light radius along the major axis, and $\epsilon$ is ellipticity.}. Discrepancies in $R_{1/2}$ ranging from 13.8 to 23.8\Arcmin persist across different studies and recent results suggest an intermediate value of $\sim$18\Arcmin. Although in principle the most recent observations cover wider area and reach deeper than older ones, there is no conclusive argument about the origin of the discrepancies. Therefore we calculate the average $R_{1/2}$ and its uncertainty using previous measurements marked with blue squares in Fig.~\ref{fig:obsHLR-SigmaLOS}. Each measurement is assumed a Gaussian distribution with its mean value and standard error and the final result is the global mean and standard deviation of these measurements. We note that we do not include the exclusively small value provided in \citet{munoz_megacam_2018}, which is not consistent with any other studies. Our result shown by the blue star gives $R_{1/2}=20.3\pm2.5$\Arcmin, which is intermediate among listed measurements.

Similarly shown in the right panel of Fig.~\ref{fig:obsHLR-SigmaLOS}, the velocity dispersion $\sigma_{\rm los}$ exhibits uncertainties of approximately $\sim$2\kms (\citealp{kleyna_photometrically_2004,walker_kinematic_2006,walker_clean_2009,wolf_accurate_2010};\citetalias{battaglia_study_2011};\citealp{tolstoy_3d_2025}). Two typical values of $7.1\pm0.3$\kms \citep{wolf_accurate_2010} and $8.4\pm0.4$\kms \citepalias{battaglia_study_2011} are given in literatures. Our work adopts the latter for that they derived the most extended LOS velocity dispersion profile for Sextans, out to a projected radius of 1.6\degree. Besides, up-to-date analysis from \citet{tolstoy_3d_2025} indicates that there are two distinct populations of stars in Sextans, metal-rich (MR) and metal-poor (MP) stars, with the latter kinematically hotter than the former. The global LOS velocity dispersion in their study is $8.3\pm0.3$\kms, which is also consistent with \citetalias{battaglia_study_2011}.

For simplicity, we assume a stellar mass-to-light ratio of 1 to derive the stellar mass $m_{*}$ of Sextans. The $V$-band luminosity, taken from \citet{wolf_accurate_2010}, yields a stellar mass of $5.9\times10^5$\Msun. We note that the adopted value is consistent with estimates in \citet{woo_scaling_2008} and \citet{karlsson_chemical_2012}. The latter approximates a stellar mass of $8.9\pm4.1\times10^5$\Msun, based on the fact that majority of the stars in Sextans are believed to be older than 10 Gyr \citep{lee_star_2009}.

The adopted parameters in this work are summarized in Table~\ref{tab:SextansObs}. In our tailored simulations, the simulated results will approach only close to the maximum-probability values. The implications of uncertainties of $R_{1/2}$, $\sigma_{\rm los}$ and $m_{*}$ are explored in Section~\ref{sec:Uncertainty in observations}.

\subsection{The satellite galaxy model}
\label{sec:The satellite galaxy model}

\subsubsection{Stellar component}
\label{sec:Stellar component}
It has been shown that stellar disk is stripped more efficiently compared with spheroidal component \citep{chang_tidal_2013} and thus can form distinct tidal structure. Since there is currently no evidence of tidal structures around Sextans, which suggests that the stellar disk of this galaxy is negligible. For simplicity, we model the stellar component of Sextans using the Plummer profile, given by:
\begin{equation}
    \rho_*(r) = \frac{3m_{*,0}}{4\pi a_*^3}\frac{1}{[1+(r/a_*)^2]^{5/2}}
    \label{eq:densstellar}
\end{equation}
where $m_{*,0}$ is the initial stellar mass and $a_*$ is the characteristic radius of this profile, corresponding to the half-mass radius of the 2D projected distribution.

\subsubsection{Dark matter halo}
\label{sec:Dark matter halo}
It is well established that the structure of DM haloes in collisionless simulations is self-similar and is characterized by a two-parameter density profile \citep[hereafter NFW]{navarro_structure_1996,navarro_universal_1997}:
\begin{equation}
    \rho_{\rm NFW}(r) = \frac{m_{\rm 200}}{4\pi[\ln{(1+c)}-c/(1+c)]}\frac{1}{r(r+r_{\rm s})^2}
    \label{eq:NFW-profile}
\end{equation}
where the profile is fully specified by the combination of the virial mass $m_{\rm 200}$ and concentration $c=r_{200}/r_{\rm s}$. The profile predicts that DM density increases as $r^{-1}$ towards the halo centre.

In order to account for baryonic effects in modifying the density structure of the progenitor DM halo of Sextans, we adopt the so-called ($\alpha,\beta,\gamma$) double power-law model \citep[][hereafter the $\alpha\beta\gamma$-model]{zhao1996},
\begin{equation}
    \rho_{\rm h}(r) = \frac{\rho_{\rm s}}{(r/a_{\rm s})^{\gamma}[1+(r/a_{\rm s})^{\alpha}]^{(\beta-\gamma)/\alpha}}
    \label{eq:dens-halo-profile}
\end{equation}
where the inner profile scales as $\sim r^{-\gamma}$. As described in \citetalias{di_cintio_mass-dependent_2014}, the parameters ($\alpha,\beta,\gamma$) are determined by the stellar-to-halo mass ratio $m_{*,0}/m_{\rm h}$ (see Eq.~3 in \citetalias{di_cintio_mass-dependent_2014}). To compute the scale density $\rho_{\rm s}$ and scale radius $a_{\rm s}$, we follow the procedure outlined in the Appendix of \citetalias{di_cintio_mass-dependent_2014}. This approach assumes that the total DM mass within $r_{200}$ (denoted as $m_{\rm h}=m_{200}-m_{*,0}$) remains unchanged by baryonic effects, leading to:
\begin{equation}
    \rho_{\rm s} = m_{\rm h}/\left[ 4\pi \int_0^{r_{200}}{\frac{r^2}{(r/a_{\rm s})^{\gamma}[1+(r/a_{\rm s})^{\alpha}]^{(\beta-\gamma)/\alpha}}\,{\rm d}r} \right]
    \label{eq:dens-halo-profile-rhos}
\end{equation}
The concentration-mass-redshift relation (CMRR) required for calculating $a_{\rm s}$ in this procedure is adopted from \citet[DM14]{dutton_cold_2014}. In the following discussion, we will refer to this relation as the CMRR unless stated otherwise. Here we note that the baryonic effect is considered only at infall, and the later evolution of the satellite after infall is only from the tidal effect of the host halo.

\subsubsection{$N$-body realization of Sextans}
\label{sec:N-body realization of Sextans}
Following the satellite model outlined in this section, we specify the parameters $m_{*,0}$, $a_*$, $m_{200}$ at the redshift $z_{\rm inf}$ when Sextans fell into the MW (derived in Section~\ref{sec:Orbital integration for Sextans}). For a given set of values \{$m_{*,0}$, $a_*$, $m_{200}$\}, we set the stellar profile using Eq.~\ref{eq:densstellar}. For the DM halo profile, we apply the procedure detailed in Section~\ref{sec:Dark matter halo} to implement the $\alpha\beta\gamma$-model from \citetalias{di_cintio_mass-dependent_2014}, which incorporates baryonic effects on the halo profile.

Sextans is modeled as a spherical isotropic two-component $N$-body system using \textsc{AGAMA} \citep{vasiliev_agama_2019}. Specifically, we use their \textsc{Spheroid} model to represent the DM component and \textsc{Plummer} model to represent the stellar component of Sextans. There is an exponential cut-off in the \textsc{Spheroid} density model. We set the cut-off radius as $1.5r_{200}$ and cut-off strength as 4. We have checked that the resulted density realization is as expected by the $\alpha\beta\gamma$-model in Eq.~\ref{eq:dens-halo-profile} within the virial radius $r_{200}$, while with a total mass of the system approximately \about$1.2$ times the virial mass $m_{200}$.

\subsection{The Milky Way model}
\label{sec:The Milky Way model}
\subsubsection{The Fiducial Milky Way model at $z=0$}
\begin{table*}
	\centering
	\caption{The parameters of our MW models at $z=0$, labelled as the Light, Fiducial and Heavy MW model.}
	\label{tab:mw3pot-static-param}
	\begin{tabular}{cccccccccc}
		\hline
        MW model & $M_{\rm 200}$& $c_{\rm MW}$& $M_{\rm d}$ & $a_{\rm d}$ & $b_{\rm d}$ & $M_{\rm b}$ & $r_{\rm b}$ & $f_{\rm d}$ & $f_{\rm b}$ \\
        & $(\rm 10^{12}\,M_\odot)$ & & $(10^{10}\,\rm M_\odot)$ & $(\rm kpc)$ & $(\rm kpc)$ & $(10^9\,\rm M_\odot)$ & $(\rm kpc)$ & (\%) & (\%) \\
        \hline
        Fiducial (M, Medium) & $1.39$ & $11.8$ & $7.93$ & $3.56$ & $0.81$ & $9.23$ & $0.22$ & 5.70 & 0.66 \\
		\hline
        Light (L) & $0.80$ & $8.5$ & $7.93$ & $3.56$ & $0.81$ & $9.23$ & $0.22$ & 9.91 & 1.15 \\
		\hline
        Heavy (H) & $2.00$ & $7.8$ & $7.93$ & $3.56$ & $0.81$ & $9.23$ & $0.22$ & 3.96 & 0.46 \\
		\hline
	\end{tabular}
\end{table*}
\begin{figure}
    \includegraphics[width=\columnwidth]{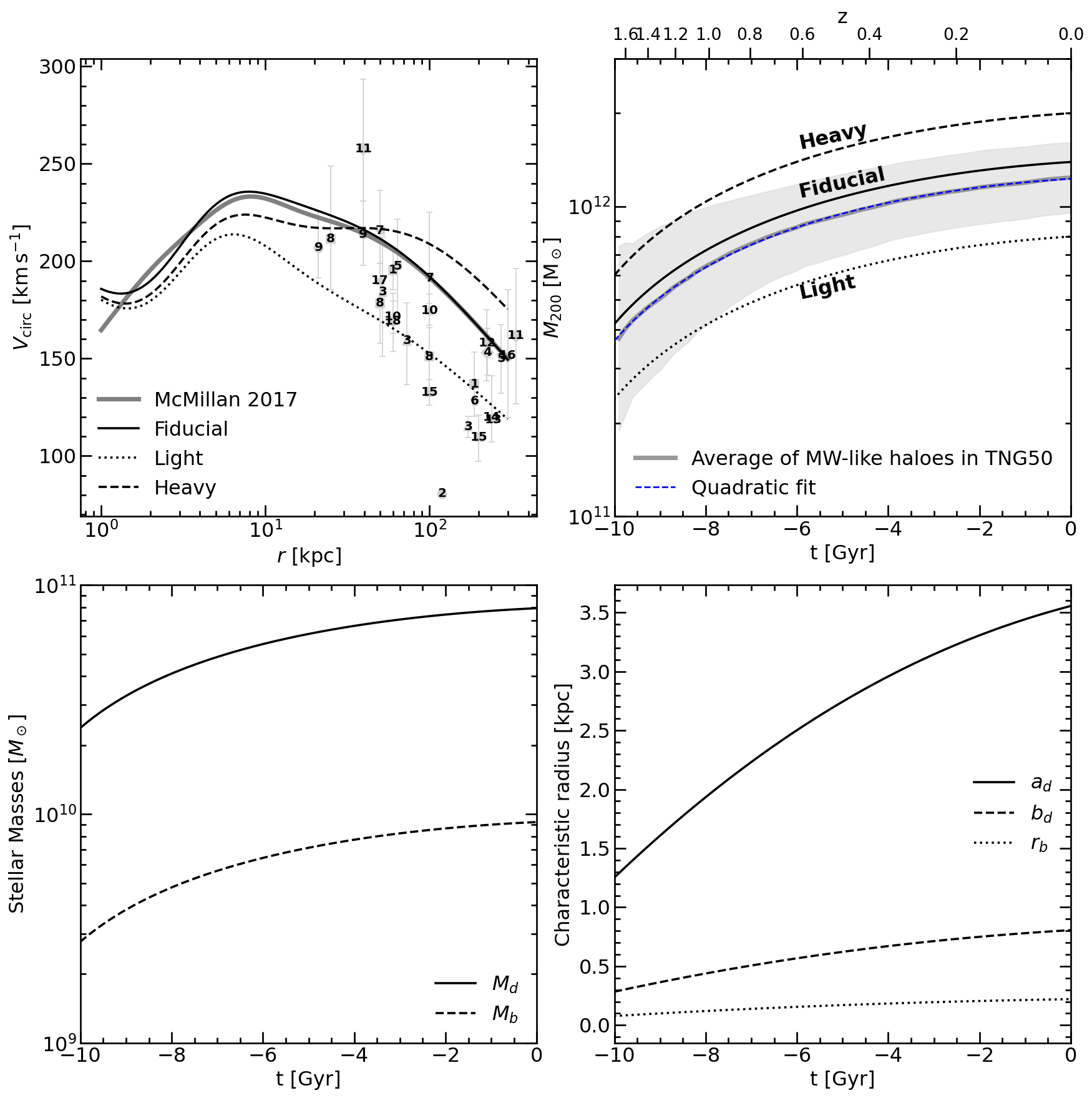}
    \caption{
    In the top-left panel, we show the circular velocity curves of our three MW models at $z=0$. The solid, dotted and dashed black lines represent the Fiducial, Light and Heavy MW model, respectively. The best-fitting model in \citetalias{mcmillan_mass_2017} are also shown by the gray line for comparison. Studies that constrain the MW mass at larger radii are shown by the gray dots with error bars (1:\citealp{zhang2025}; 2:\citealp{ou2024}; 3:\citealp{bird_milky_2022}; 4:\citealp{liConstrainingMilkyWay2020}; 5:\citealp{fritzMassOurGalaxy2020}; 6:\citealp{eilersCircularVelocityCurve2019}; 7:\citealp{vasiliev2019}; 8:\citealp{eadie2019}; 9:\citealp{watkinsEvidenceIntermediatemassMilky2019}; 10:\citealp{fardal2019}; 11:\citealp{sohnAbsoluteHubbleSpace2018}; 12:\citealp{mcmillan_mass_2017}; 13:\citealp{bovy_galpy_2015}; 14:\citealp{kafle2014}; 15:\citealp{gibbons2014}; 16:\citealp{boylan-kolchin2013}; 17:\citealp{deason2012}; 18:\citealp{xue2008}).
    In the rest of panels, we show the evolution of the MW parameters as a function of time. The top-right panel shows the virial mass evolution. The thick gray line shows the averaged halo mass evolution of the MW-like haloes in TNG50 and the shaded region represents the $1\sigma$ deviation. The averaged halo mass evolution is well approximated by a quadratic function shown by the blue dashed line. The lower panels show evolution of parameters of stellar disk and stellar bulge.
    }
    \label{fig:mw-params-evo}
\end{figure}
Our MW model consists of three components: an axisymmetric Miyamoto-Nagai stellar disk \citep{miyamoto_three-dimensional_1975}, a Hernquist stellar bulge \citep{hernquist_analytical_1990}, and a NFW DM halo \citep{navarro_structure_1996, navarro_universal_1997}. The gravitational potential of these models can be expressed analytically as:
\begin{equation}
    \begin{aligned}
        &\phi_{\rm d}(R,z) = -\frac{GM_{\rm d}}{\sqrt{R^2+\left(a_{\rm d}+\sqrt{z^2+b_{\rm d}^2}\right)^2}}\\
        &\phi_{\rm b}(r) = -\frac{GM_{\rm b}}{r+r_{\rm b}}\\
        &\phi_{\rm h}(r) = -\frac{G(M_{\rm 200}-M_{\rm d}-M_{\rm b})}{\ln{(1+c_{\rm MW})}-c_{\rm MW} / (1+c_{\rm MW})}\frac{\ln{(1+r / r_{\rm MW,s})}}{r}
        \label{eq:mw3pot-static}
    \end{aligned}
\end{equation}
where the model is defined by 8 parameters: $M_{\rm d}$, $a_{\rm d}$ and $b_{\rm d}$ represent the mass, characteristic radius, and scale height of the stellar disk, respectively; $M_{\rm b}$ and $r_{\rm b}$ denote the mass and characteristic radius of the stellar bulge; $M_{\rm 200}$ is the virial mass of the MW, $r_{\rm MW,s}$ is the scale radius and $c_{\rm MW}\equiv r_{\rm MW,200}/r_{\rm MW,s}$ is the concentration parameter, where $r_{\rm MW,200}$ is the virial radius.

Observational constraints on these eight parameters remain modestly uncertain. For instance, the mass of the stellar disk is estimated to lie within the range 3\about10$\times10^{10}$\Msun (e.g., \citealt{mcmillan_mass_2011,bovy_galpy_2015,battaglia_effect_2015,mcmillan_mass_2017,pouliasis_milky_2017}, see \citealt{bland-hawthorn_galaxy_2016} for a review). In our Fiducial model, we adopt disk, bulge and halo masses and scale radii similar to those in \citet[M17]{mcmillan_mass_2017}, with minor adjustments to better match the potential given in their model. We note that the concentration $c_{\rm MW}$ in this model deviates from the CMRR by $+0.165$\dex, following the best-fitting model of \citetalias{mcmillan_mass_2017}.

In the top-left panel of Fig.~\ref{fig:mw-params-evo}, the solid black line and gray line show the circular velocity curve in a range of 1-300\kpc of our Fiducial MW model and the best-fitting \citetalias{mcmillan_mass_2017} model respectively. Our model matches \citetalias{mcmillan_mass_2017} pretty well in the outer region at radius larger than \about60\kpc, while with minor difference in the inner region. Such difference has negligible influence the tidal history of Sextans, since it never enters the inner part of our MW (see Section~\ref{sec:Orbital integration for Sextans}).

The parameters of our Fiducial model are listed in Table~\ref{tab:mw3pot-static-param} for reference.

\subsubsection{The uncertainty of the Milky Way mass}
The mass of the MW is better constrained in the inner \about20\kpc with good quality data for either circular velocity profile or Jeans modelling, while the poor data at larger distance leads to a total MW mass within a range of 0.5\about2$\times10^{12}$\Msun (see the reviews by \citealt{wang_mass_2020} and \citealt{bobylev_modern_2023}). To investigate the impact of the MW's mass on the tidal evolution of Sextans, we introduce two variant MW models: the Light (L) and Heavy (H) model. The Light model assumes a virial mass of $8\times10^{11}$\Msun and the Heavy model assumes $2\times10^{12}$\Msun. Unlike our Fiducial model, the concentration of DM haloes in the two models are derived from the CMRR. In both models, the parameters of the stellar components are held constant and thus only the halo parameters are altered from the Fiducial model.

In the top left panel of Fig.~\ref{fig:mw-params-evo}, we add more recent observational constraints on the outer circular velocity profile of the MW. It is seen that a wide range of circular velocities are allowed by current data constraints, and our adopted Light and Heavy models are within the data constraints. Note that there is disagreement between the Light/Heavy model with \citetalias{mcmillan_mass_2017} at $r<50$\kpc, simply because we use an average CMRR from simulation for DM halo. We tested that we could better fit the profile of \citetalias{mcmillan_mass_2017} in the inner region by changing the halo concentration and it has negligible effect on the evolution as Sextans's orbit is largely determined by the outer halo mass and its pericenter is larger than 50\kpc. The detailed parameters for these MW models at $z=0$ are listed in Table~\ref{tab:mw3pot-static-param}.

\subsubsection{The evolution of the Milky Way}
\label{sec:The evolution of the Milky Way}
In the \LCDM paradigm, DM haloes grow through the accretion of diffuse mass and the mergers of smaller dynamical objects, implying that the mass of the MW was smaller in the past. In this section we present an analytic method to model the mass evolution of the MW. The evolution of the MW is also used to constrain the infall time of the satellite galaxy.

The growth of halo mass in \LCDM model has been extensively studied using simulations \citep[e.g.,][]{wechsler_concentrations_2002,zhao_growth_2003,genel_mergers_2008,mcbride_mass_2009,fakhouri_merger_2010}. As we will compare the recovered infall mass of Sextans with the TNG50 simulation, we use the TNG50 Milky Way+Andromeda Sample \citep{pillepich_milky_2024} to model the evolution of our MW. The halo mass $M_{200c}$ at $z=0$ is confined within $0.8$\about$2\times10^{12}$\Msun, resulting in total 117 MW-like haloes. We compute the average mass of these MW-like haloes over time that is shown by the gray line in the top-right panel of Fig.~\ref{fig:mw-params-evo}. The shaded region in this panel represents the $1\sigma$ scatter. The averaged mass evolution of MW-like haloes is very smooth, with only $\sim4\times10^{11}$\Msun 10\Gyr ago and $1.23\times10^{12}$\Msun at $z=0$, which is close to our Fiducial MW. As is shown by the blue line in this panel, the averaged mass evolution can be well approximated by a quadratic function, which is
\begin{equation}
\frac{M(t)}{[10^{10}{M_\odot}]} = 123 + 2.60 \cdot \left( \frac{t}{[\rm Gyr]} \right) - 0.600 \cdot \left( \frac{t}{[\rm Gyr]} \right)^2
\label{eq:mwacc-mass-TNG50}
\end{equation}
where $M(t)$ represents the halo mass at time $t$. Here, $t$ is the negative lookback time, with $t=0$ corresponding to the present time (or $z=0$). 

The evolution of virial mass in our three MW models is then scaled by
\begin{equation}
    \begin{aligned}
        M_{\rm 200, k}(t) = M_{\rm 200, k}(t=0)\cdot \frac{M(t)}{M(t=0)}
        \label{eq:mwacc-mass}
    \end{aligned}
\end{equation}
where $M_{\rm 200,k}$ is the virial mass for the k-th MW model (k={Fiducial, Light, Heavy}). The concentration of DM halo of the MW is also evolved, following the CMRR from \citetalias{dutton_cold_2014}. In the Light and Heavy MW model, it follows CMRR explicitly, while in the Fiducial MW model, it deviates by $+0.165$\dex from the CMRR expectation at all redshifts due to the deviation at $z=0$.

The evolution of the stellar components of the host halo is more uncertain compared to the DM component. For simplicity, we fix the mass ratio (shown in Table~\ref{tab:mw3pot-static-param}, $f_{\rm d}$ for stellar disk and $f_{\rm b}$ for bulge) between the stellar and DM components over time. The characteristic radii are scaled proportionally to the MW virial radius, $f_{\rm r} = r_{\rm MW,200}(t)/r_{\rm MW,200}(t=0)$.

In Fig.~\ref{fig:mw-params-evo}, we show the evolution of MW masses and scale radii. The mass evolution in our models aligns well with that of TNG50. The masses in different components evolve about 3.3 times and the scale radii evolve 2.8 times within 10\Gyr. We note that the MW mass evolution in our models approximates only the average evolution in cosmological simulations, where the massive mergers or perturbations in the accretion history of our MW are not accounted for.

\subsection{Orbital integration for Sextans}
\label{sec:Orbital integration for Sextans}
\begin{table*}
    \centering
    \caption{The initial positions and velocities of the orbits obtained by our OI for the 9 simulations. The second and third columns are the positions and velocities at time $t_{\rm IC}$. The forth and fifth columns are the infall time and the corresponding redshift. The last column is the time $t_{\rm IC}$ when we put the satellite at the start of the simulation.}
    \label{tab:icPS}
    \begin{tabular}{cccccc}
    \hline
    Orbit model & (x, y, z) & ($v_{\rm x}$, $v_{\rm y}$, $v_{\rm z}$) & $t_{\rm inf}$ & $z_{\rm inf}$ & $t_{\rm IC}$ \\
     & (kpc) & ($\rm km\,s^{-1}$) & (Gyr) & & (Gyr) \\
    \hline
    L81   & (62.00, -174.50, 67.20) & (-107.30, 73.96, 23.20) & -7.92 & 1.00 & -8.92 \\
    L86   & (233.65, 56.68, -166.33) & (-38.13, -77.03, 72.72) & -1.21 & 0.09 & -2.21 \\
    L91   & (275.69, 7.38, -162.02) & (-80.20, -65.73, 89.25) & -1.06 & 0.08 & -2.06 \\
    \hline
    M74    & (47.42, -159.54, 68.38)  & (-94.10, 11.00, 52.52) & -7.38 & 0.87 & -8.38 \\   
    M81    & (-93.01, -111.73, 126.24) & (-50.53, 112.98, -47.21) & -8.54 & 1.16 & -9.54 \\   
    M88    & (207.36, -98.36, -22.15) & (-57.56, -57.30, 68.44) & -5.62 & 0.56 & -6.62 \\
    \hline
    H70    & (-131.89, 50.89, 44.13)  & (14.96, 104.91, -76.38) & -8.27 & 1.09 & -8.77 \\  
    H78    & (-141.76, 112.86, 4.45)  & (86.39, 45.52, -78.62) & -9.03 & 1.32 & -10.03 \\ 
    H85    & (-72.51, -135.23, 131.41) & (-95.98, 71.15, 0.16) & -7.45 & 0.89 & -8.45 \\
    \hline
    \end{tabular}
\end{table*}
\begin{figure*}
    \includegraphics[width=2\columnwidth]{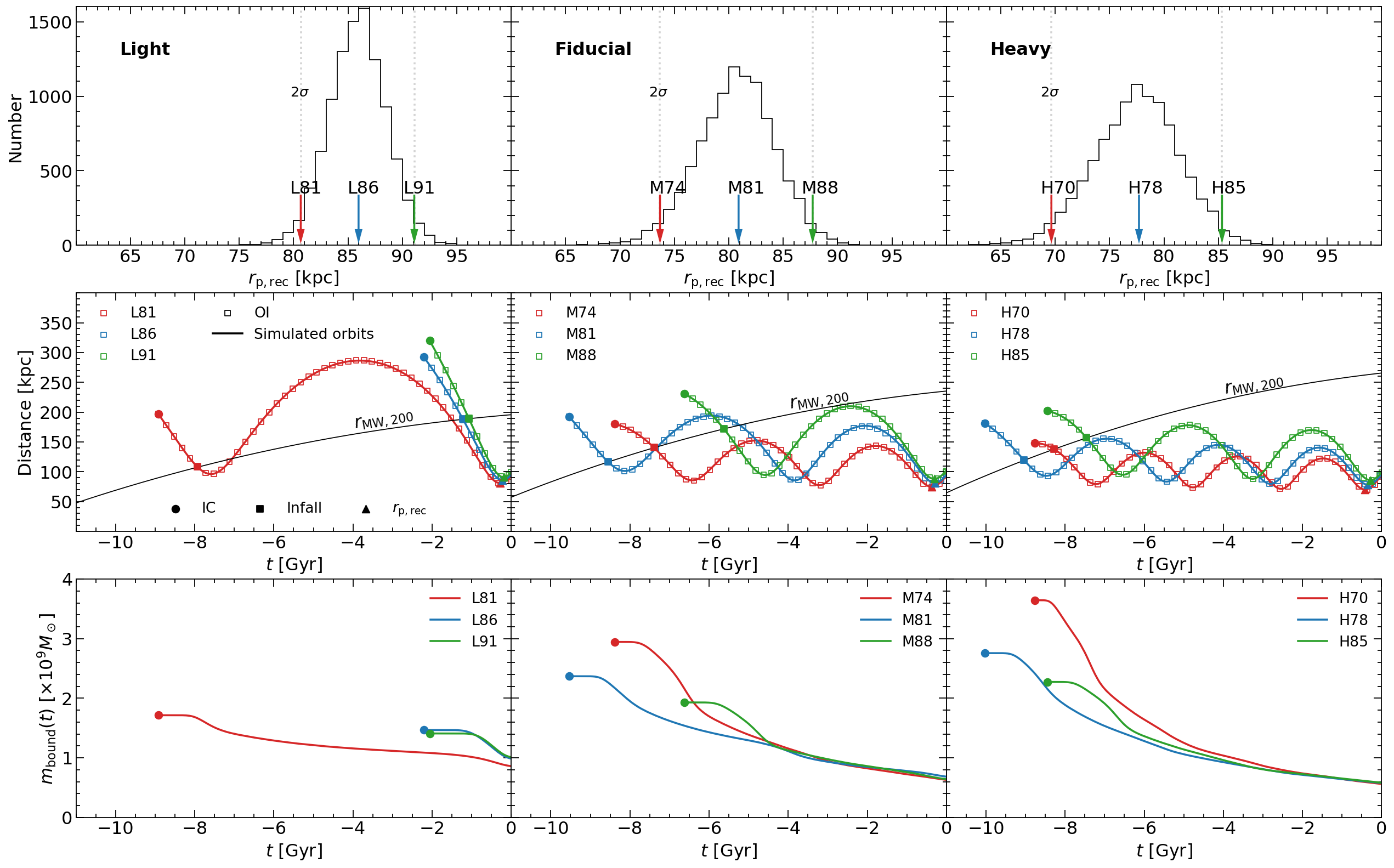}
    \caption{
    {\bf Top}: the distribution of $r_{\rm p,rec}$ of the 10,000 Monte Carlo realizations of orbits for Sextans in three MW models. The downward arrows denote the $r_{\rm p,rec}$ of three representative orbits chosen for simulations. Blue represents orbits with medium pericenter distance, while red and green represent orbits with smaller and larger values, which are $2\sigma$ away from the medium value. As is denoted by the text, we name these orbits as L81, L86, L91, M74, M81, M88, H70, H78 and H85 according to the MW model and $r_{\rm p,rec}$.
    {\bf Middle}: the specifically chosen orbits that have different pericenter distance. The black lines mark the evolution of virial radius $r_{\rm MW,200}$ of the host MW. Squares mark the point of infall and upper triangles mark the most recent pericenter. The solid circles denote the start point of simulations.
    {\bf Bottom}: the bound DM mass evolution over time of the 9 simulations. Colors are marked according to the upper conventions.
    }
    \label{fig:OI-Sextans}
\end{figure*}
After constructing the MW model, the orbit of a MW satellite galaxy can be immediately backward integrated from its present center-of-mass (COM) position and velocity. Thanks to the \textit{Gaia} mission, we now have access to more precise PMs of MW satellite galaxies. In this work, we adopt the PMs for Sextans from \citet{li_gaia_2021}, which are up-to-date results of \textit{Gaia} EDR3. The other data, such as sky positions, distance and LOS velocity, are summarized in Table~\ref{tab:SextansObs}.

We do not include dynamical friction \citep{chandrasekhar_dynamical_1943} when integrating the orbits of Sextans. The dynamical friction exerted on Sextans is expected to be secondary since the mass ratio between Sextans and the MW is at the order of $10^{-3}$. We have tried including the dynamical friction assuming the recovered infall masses in this work using \textsc{galpy} \citep{bovy_galpy_2015} and it is found that the maximum orbital decay caused by dynamical friction is smaller than 5 per cent. For self-friction, another friction force decaying the orbits, it's dependent on the stripped mass of a subhalo and is even secondary over dynamical friction \citep{miller_dynamical_2020}. We thus do not include self-friction either when conducting orbital analysis here. However we adjust the integrated orbits by incorporating a friction term to ensure best match with simulated ones when conducting our tailored simulations (see Section~\ref{sec:Simulation set-up}).

The orbital integration (OI) is conducted in a right-handed galactocentric coordinate system. We adopt the Sun parameters used in \citetalias{mcmillan_mass_2017}, specifically the position $(R_\odot, z_\odot)=(8.21, 0.025)$\kpc, a circular velocity of $233.1$\kms and the Local Standard of Rest velocity $(U_\odot,V_\odot,W_\odot)=(11.1,12.24,7.25)$\kms.

To account for the statistical uncertainties in the distance and velocity of Sextans, we perform 10,000 Monte Carlo realizations of its possible orbits, assuming a Gaussian distribution centered on the mean values of the observed distance and velocities. The top panels of Fig.~\ref{fig:OI-Sextans} show the distribution of the $r_{\rm p,rec}$, which is the distance from galactic center to the most recent pericenter, for these orbits across different MW models. The $r_{\rm p,rec}$ values span a wide range, from 65 to 95\kpc, depending on the MW model. For each MW mass model, we select three representative orbits with small, medium, and large pericenter distances for Sextans. The small- and large-pericenter orbits are $2\sigma$ away from the peak, meaning the probability of having an orbit closer or farther than the selected one is only about 2 per cent. These selected orbits are indicated by downward arrows in the plot, which are named according to the MW model and their $r_{\rm p,rec}$ values.

The middle panels of Fig.~\ref{fig:OI-Sextans} illustrate these chosen orbits, with squares representing the OI results and solid lines showing the actual orbits extracted from the simulations. In each panel, green, blue, and red lines denote orbits with large, medium, and small pericenter distances, respectively, while the black line represents the virial radius of the host MW as a function of time. As shown, the pericenter distances of these orbits decay over time primarily due to the mass growth of the MW. By this way, the infall time is defined as the time satellite firstly cross the virial radius of the MW.

Sextans has passed through the pericenter for $2$\about$3$ times since its infall in our Fiducial MW model, $1$\about$2$ times in the Light MW model, and $3$\about$4$ times in the Heavy MW model. Consequently, Sextans is expected to experience different tidal stripping processes in these runs, as is shown by the bound mass evolution in the bottom panels of Fig.~\ref{fig:OI-Sextans}. Detailed analysis will be explored in the results section.

We conduct our simulation runs by placing Sextans on these chosen orbits to study its dynamical evolution. The initial COM positions and velocities of Sextans are values at the time $t_{\rm IC}$, typically 1\Gyr (except for H70) earlier than the infall time $t_{\rm inf}$. These values are listed in Table~\ref{tab:icPS} for each simulation run. Hereafter, we will refer to the simulation runs using the orbit model names.

\subsection{Simulation set-up}
\label{sec:Simulation set-up}
To simulate a galaxy that closely matches the observed properties of Sextans, we employ a tailored simulation technique similar to those used for Fornax \citep{battaglia_effect_2015} and Sculptor \citep{iorio_effect_2019}. This approach allows us to determine the optimal parameters for the progenitor of Sextans, such as its infall virial mass $m_{\rm 200}$ and inner density profile.

For the i-th simulation, we firstly construct a MW model and perform OI for Sextans to determine its initial position $\vec{r}^{(i)}$ and velocity $\vec{v}^{(i)}$ at the time $t_{\rm IC}^{(i)}$, and the infall time $t_{\rm inf}^{(i)}$ and redshift $z_{\rm inf}^{(i)}$. Then we adopt an iterative method to conduct our simulation:
\begin{itemize}
    \item[(i)] Initial parameter estimation for Sextans: A first guess of $m_{*,0}$, $a_*$ is simply its observed stellar mass and half-light radius, as Sextans is not expected to have undergone significant tidal stripping of its stellar component \citep[e.g.,][]{cicuendez_tracing_2018}. The initial guess of $m_{200}$ for Sextans is assumed to be $10^{9}M_{\odot}$ for all simulations;
    \item[(ii)] $N$-body realization: We generate a satellite galaxy with both stellar and DM halo components using \textsc{AGAMA} \citep{vasiliev_agama_2019}. The system is evolved in isolation for 1\Gyr using the publicly available \textsc{GADGET-4} simulation code \citep{springel_simulating_2021} to ensure stability;
    \item[(iii)] The main simulation: We set our analytic MW model as an external gravity field in \textsc{GADGET-4}, and the satellite galaxy is placed at $\vec{r}^{(i)}$ with COM velocity of $\vec{v}^{(i)}$. The modified \textsc{GADGET-4} code is then used to evolve the system from $t_{\rm IC}^{(i)}$ to the present time ($t=0$);
    \item[(iv)] Calculation of present-day properties: The center and velocity of the satellite galaxy are calculated using stellar particles, following the scheme outlined in Appendix A of \citet{van_den_bosch_disruption_2018}. We also calculate the bound stellar mass $m_*^{\rm sim}$, the half-light radius $R_{1/2}^{\rm sim}$ and the LOS velocity dispersion $\sigma_{\rm los}^{\rm sim}$ within a projected radius of $1.6^\circ$ for comparison with observations;
    \item[(v)] Parameter updates and iteration: The observational data for Sextans (Table~\ref{tab:SextansObs}) serve as constraints. If the simulation results fall within the observational error margins, the procedure is terminated. Otherwise, the initial parameters are updated using the following estimator:
    \begin{equation}
        \begin{aligned}
            &m_{200}^{(i+1)} = m_{200}^{(i)}\cdot\left(\frac{\sigma_{\rm los}}{\sigma_{\rm los}^{\rm sim}}\right)^2 \\
            &m_{*,0}^{(i+1)} = m_{*,0}^{(i)}\cdot\frac{m_*}{m_*^{\rm sim}} \\
            &a_*^{(i+1)} = a_*^{(i)}\cdot\frac{R_{1/2}}{R_{1/2}^{\rm sim}} \\
            \label{eq:update-param}
        \end{aligned}
    \end{equation}
    With these updated parameters, the DM density slope is again determined by the \citetalias{di_cintio_mass-dependent_2014} relation. We note that the simulated orbit does not match the OI perfectly due to the self-friction force exerted by the tidally stripped mass of Sextans \citep[e.g.][]{miller_dynamical_2020}. To ensure the integrated orbit best matches the simulated one, we adjust OI by incorporating an empirical friction term, namely $\vec{f}_{\rm SF}=-A\frac{\vec{v}}{v^B}$, where $v$ is the norm of velocity $\vec{v}$. The parameter $A$ and $B$ for the next simulation is fitted by the simulated orbit in the i-th iteration. Then the $\vec{r}^{(i+1)}$, $\vec{v}^{(i+1)}$, $t_{\rm IC}^{(i+1)}$, $t_{\rm inf}^{(i+1)}$ and $z_{\rm inf}^{(i+1)}$ are obtained by the adjusted orbit. The process then repeats from step (ii) until convergence is achieved.
\end{itemize}
By following the above procedures, we can finally obtain the optimal values of $m_{*,0}$, $a_*$ and $m_{200}$ for Sextans at its infall typically after 3\about5 iterations, ensuring the best match to its present-day properties.

For all of simulation runs, we set the number of DM particles as $N_{\rm h}=10^7$. The softening radius for DM particles $\epsilon_{\rm h}\approx25$\pc is calculated using Eq.~15 of \citet{power_inner_2003}, ensuring enough spatial resolution to capture the evolution of kinematics within the half-light radius of Sextans. For the stellar component, we fix the number of particles at $N_*=5\times10^4$ and the softening radius at $\epsilon_*=10$\pc. The time interval between two close snapshots is set to be 10$\,$Myr, which is small enough to resolve the dynamical evolution of Sextans around the pericenter. We verified the numerical convergence and no spurious heating of our results by performing additional simulations with variations in the number of particles and softening lengths and by running the satellite galaxy in isolation for another 10\Gyr.

\section{Results}
\label{sec:Results}

\subsection{Comparison with observations}
\label{sec:Comparison with observations}
\begin{figure*}
    \includegraphics[width=2\columnwidth]{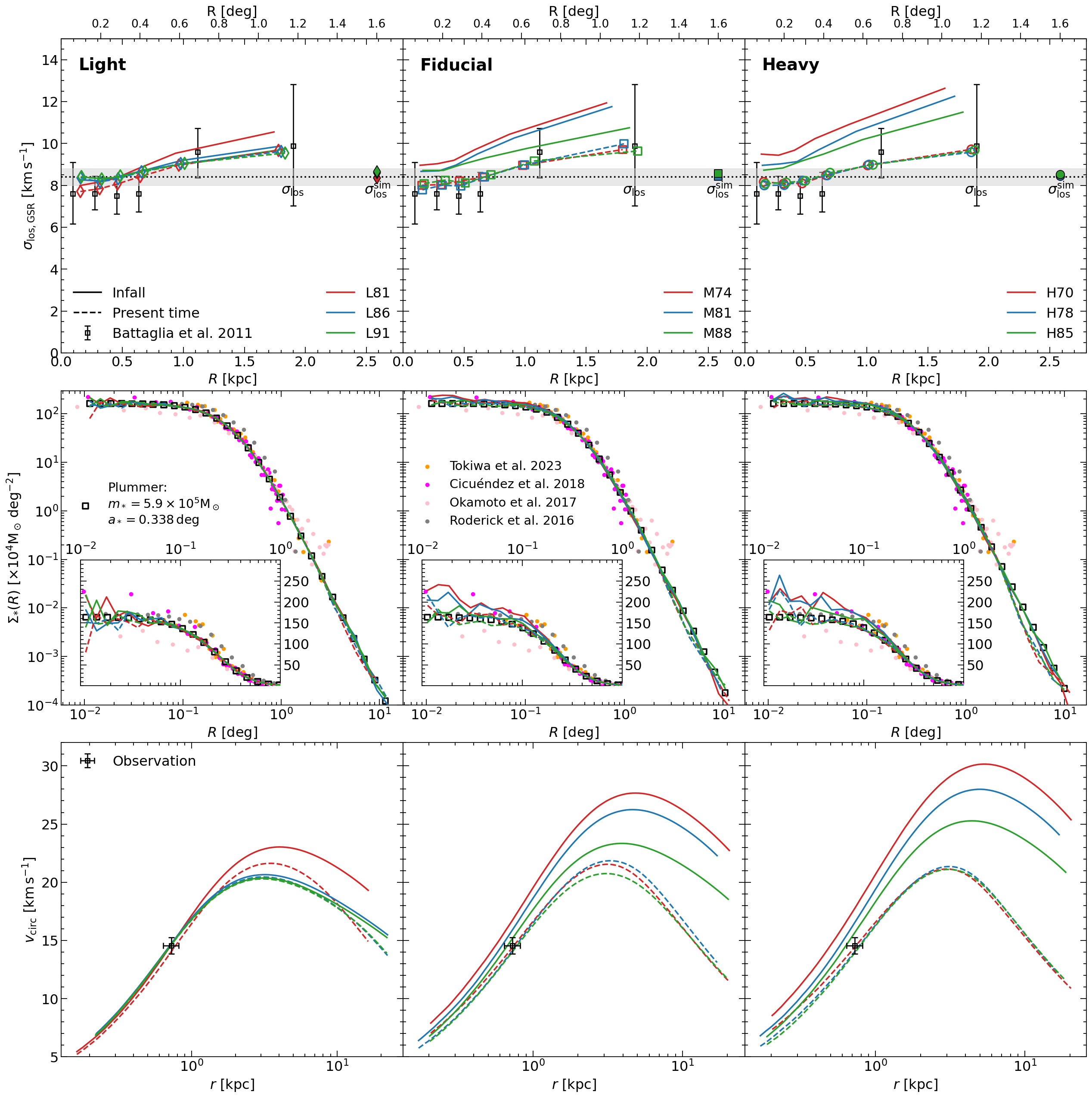}
    \caption{ 
    The simulated properties of the satellite galaxies at infall (solid lines) and present time (dashed lines) across different MW models, with red, blue and green denoting simulations respectively with close, medium, and far pericenter orbits, compared with those of observation (black squares).
    {\bf Top}: LOS velocity dispersion as a function of projected radius $R$ of stellar component in different simulations compared with observed one \citepalias{battaglia_study_2011}. See the text for details.
    {\bf Middle}: the projected surface density profile of stellar component in simulations compared with observation. The black squares are 2D Plummer profile with scale radius of half-light radius $R_{1/2}$ and total stellar mass of $m_*$ in Table~\ref{tab:SextansObs}. The insets show the zoom-in profiles within $1^\circ$. For comparison, the surface density profiles measured by a few observation work are also shown using color dots.
    {\bf Bottom}: the circular velocity of DM component as a function of radial distance $r$ in simulations. The black squares with error bars are derived circular velocity at $r_{1/2}$ as in Table~\ref{tab:SextansObs}.
    }
    \label{fig:sim-obs}
\end{figure*}
\begin{figure*}
    \includegraphics[width=2\columnwidth]{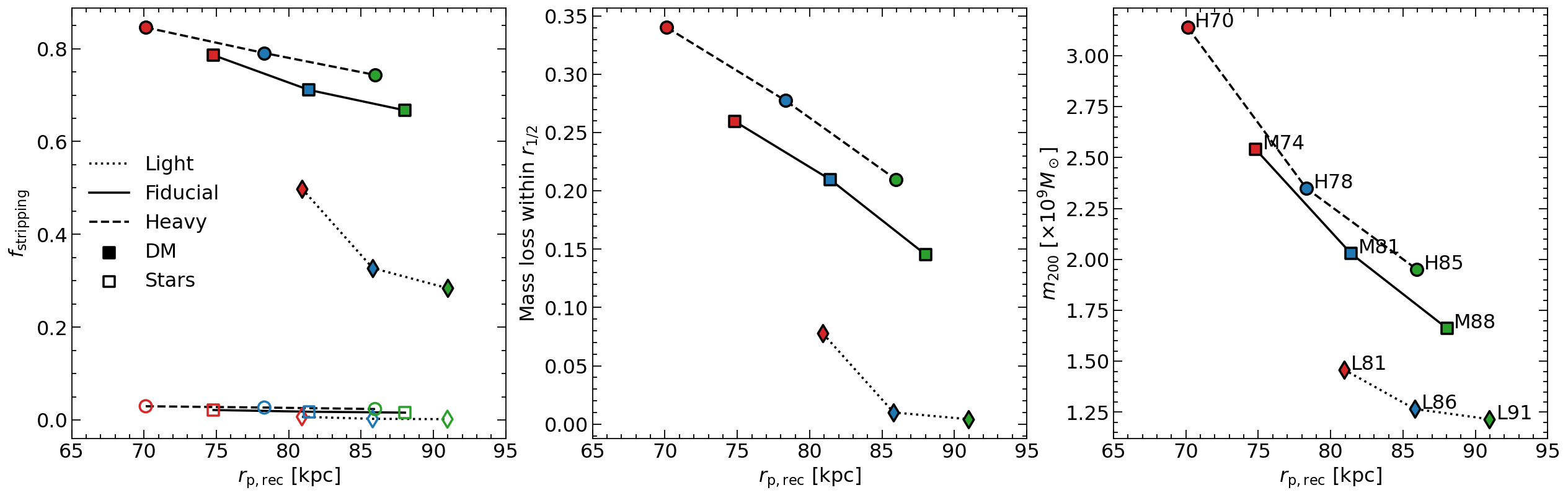}
    \caption{ 
    The dependence on the MW model and pericenter distance of the tidal history of Sextans with dotted lines for Light MW model, solid lines for Fiducial MW model and dashed lines for Heavy MW model. Symbols with different colors represent different simulations, with the simulation name labeled in the right panel. The left panel shows the bound mass loss fraction from infall to present time, defined by $f_{\rm stripping}\equiv1-m_{\rm bound}(t=0)/m_{\rm bound, inf}$, versus the the most recent pericenter distance $r_{\rm p,rec}$. The points with solid box are bound mass loss fraction for DM, while those with hollowed box are for stellar component. The middle panel shows the DM mass loss fraction within a radius of $r_{1/2}$. The right panel shows the infall virial mass of Sextans recovered by our simulations.
    }
    \label{fig:sim-tidal}
\end{figure*}
We compare our simulation results with the observed properties of Sextans in Fig.~\ref{fig:sim-obs}. In each panel, the observables are represented by black squares, while the corresponding simulated quantities at infall and present time are depicted respectively as solid and dashed lines, with red, blue and green denoting simulations respectively with close, medium, and far pericenter orbits across the three MW models.

In the top panels of Fig.~\ref{fig:sim-obs} we compare the LOS velocity dispersion profiles with those obtained by \citetalias{battaglia_study_2011}. The black squares with error bars are from Fig.~15 of \citetalias{battaglia_study_2011}, which shows a slightly increasing profile. Since \citetalias{battaglia_study_2011} converted the heliocentric LOS velocities to radial velocities with respect to Galactic center ($v_{\rm los,GSR}$, where GSR stands for Galactic standard of rest), we apply the same transformation to our simulated data using \textsc{galpy} \citep{bovy_galpy_2015}. We note that the data in \citetalias{battaglia_study_2011} are calculated with elliptical binning and we scale the radii with $\sqrt{1-\epsilon}$, where $\epsilon=0.35$ for comparison. Then the simulated values are calculated at the similar binning radii with each annulus containing 8400 particles. As shown in these panels, the simulated velocity dispersion profiles also exhibit an increasing trend with radius and all nine simulations are globally consistent with the observations, despite in the inner region the values are slightly higher while still in the error margins. The LOS velocity dispersion within $R<1.6$\degree for the stellar component is constrained to be close to $\sigma_{\rm los}=8.4$\kms, shown by the solid diamonds, squares and circles in the top panels.

As mentioned in Section~\ref{sec:The observed properties of Sextans} that the 2D Plummer profile provides a good fit to the surface number density of Sextans. The scale radius of this profile corresponds to the half-light radius. In the middle panels of Fig.~\ref{fig:sim-obs}, we thus compare our simulated stellar distribution with a 2D Plummer density distribution with a scale radius of observed $R_{1/2}$ and stellar mass of $m_*$ (as listed in Table~\ref{tab:SextansObs}). This profile out to a projected radius of \about$10^\circ$ is shown by black squares in these panels, with the insets showing the zoom-in profile within $1^\circ$. The simulated stellar distributions at present time that are calculated directly from the stellar particles are shown as dashed lines. The simulated profiles match well with the 2D Plummer profile within a projected radius of \about$1^\circ$, where the number density profiles are measured by several studies \citep[e.g.][]{roderick_structural_2016,okamoto_population_2017,cicuendez_tracing_2018,tokiwa_study_2023}. For comparison, we convert the number density profiles from these studies into surface density assuming a total stellar mass of $5.9\times10^5$\Msun, which are shown by dots in these panels. As expected, our simulated profiles are well within the range of the observed ones, since we adopt the averaged half-light radii measured in previous work.

The bottom panels of Fig.~\ref{fig:sim-obs} show the simulated circular velocity profiles of the DM component within a range from $0.01r_{200}$ to $r_{200}$ at infall and present time. The black square with error bars in each panel represents the derived circular velocity $v_{1/2}$ at the 3D de-projected half-light radius $r_{1/2}$ for Sextans, as shown in Table~\ref{tab:SextansObs}. Our simulated results are in excellent agreement with the observation across all possible orbits, indicating that stellar kinematics of Sextans provide robust constraints on dynamical mass within $r_{1/2}$, which is implemented by many works on estimating dynamical mass of MW dSphs \citep{walker_universal_2009,wolf_accurate_2010}.

\begin{figure*}
    \includegraphics[width=2.0\columnwidth]{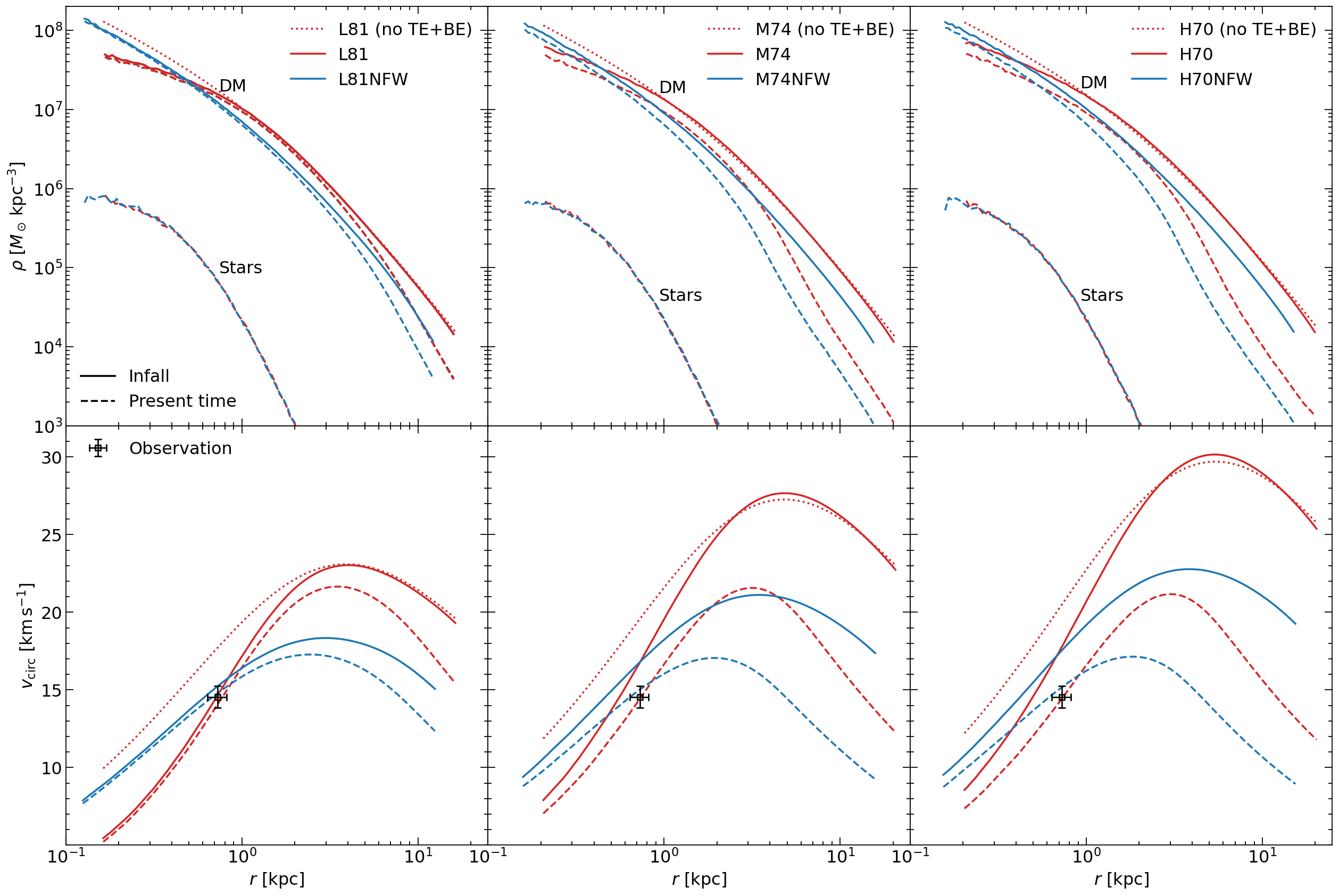}
    \caption{The density (upper panels) and circular velocity (lower panels) profiles of DM component in simulation L81, M74 and H70 (red lines), compared with their NFW counterparts L81NFW, M74NFW and H70NFW (blue lines), at infall (solid lines) and at present (dashed lines). The dotted lines denote the theoretical NFW models assuming virial masses and CMRR concentrations of L81, M74 and H70 while not considering tidal and baryonic effects. In the upper panels, we also show the present density profiles of the stellar component in these simulations.}
    \label{fig:sim-vcirc}
\end{figure*}

\subsection{Tidal evolution of Sextans}
\label{sec:Tidal evolution of Sextans}
Based on the present constraints on MW mass and PM measurements, the orbital history of Sextans exhibits large uncertainties with the pericenter distance and number of pericenter passages, as is shown in Fig.~\ref{fig:OI-Sextans}. There are fewer pericenter passages and the pericenter is further out for a lower mass MW. For example, there are only 1\about2 pericenters with an average $r_{\rm p,rec}$ of \about86\kpc for the Light MW model, 2\about3 pericenters with average $r_{\rm p, rec}$ of \about81\kpc for the Fiducial MW model, 3\about4 pericenters with average $r_{\rm p,rec}$ of \about78\kpc for the Heavy MW model. Fixing the MW potential, there is still 10\about15\kpc uncertainty for the $r_{\rm p,rec}$ at a $2\sigma$ level.

The uncertainty in orbital history causes uncertainty in mass loss of Sextans. As is shown in the bottom row of Fig.~\ref{fig:OI-Sextans}, where we show the bound DM mass evolution of our simulations, the bound DM mass evolves from $[1.4,1.5,1.7]\times10^9$ to $\sim$$10^9$\Msun in the Light MW, from $[1.9,2.4,3.0]\times10^9$ to $\sim$$6\times10^8$\Msun in the Fiducial MW and from $[2.3,2.8,3.6]\times10^9$ to $\sim$$6\times10^8$\Msun in the Heavy MW. It is interesting but not surprising to note that the final DM mass of Sextans is similar in the Fiducial and Heavy MW model as the present observational properties of Sextans are used as constraints, while in the Light MW, the final DM mass is slightly higher than that in the other two MW models, due to a later infall time and weak tidal stripping in this Light MW model.

The left panel of Fig.~\ref{fig:sim-tidal} depicts how the tidal mass loss of Sextans depends on the MW mass as well as the orbit uncertainty. It is found that the tidal stripping fraction for the DM component varies significantly across different MW models, from 32.8 to 79.0 per cent for the medium pericenter orbit, corresponding to an uncertainty of \about46 per cent for a MW mass between $0.8$\about$2\times10^{12}$\Msun. By contrast, the uncertainty over the orbital pericenter is smaller with only 10\about20 per cent uncertainty. This demonstrates that the uncertainty in MW mass dominates the uncertainty of tidal evolution for the DM component of Sextans.

However, for the stellar component, which is shown by hollowed symbols in the left panel of Fig.~\ref{fig:sim-tidal}, the maximum tidal mass loss fraction in our simulation is no more than 5 per cent for a MW mass smaller than $2\times10^{12}$\Msun, indicating the stars in Sextans are not or only mildly affected by galactic tides.

The weak tidal feature for stars is also validated by stellar distribution shown in Fig.~\ref{fig:sim-obs}, where the minor change of the surface density profile from infall to present time indicates that tides are not effective in disturbing the stellar distribution. Most data points suggest that the density distribution is not disturbed out to a radius of 1\degree. This is consistent with our simulation where the tentative signs of tidal disturbance are only seen out of a projected radius of \about$3^\circ$.

Unlike the surface density, the stellar velocity dispersion drops obviously from infall to present time in the Fiducial and Heavy MW model, as is shown in the top panels of Fig.~\ref{fig:sim-obs}. This change is mainly caused by the relaxation of stellar component due to the loss of DM mass in the inner region. It can be found in the bottom panels of Fig.~\ref{fig:sim-obs} that the DM circular velocity at $r_{1/2}$ decreases 1\about3\kms among these models. In the Light MW model, the stellar velocity dispersion is nearly unchanged due to the weak tidal stripping. In the middle panel of Fig.~\ref{fig:sim-tidal} we show the DM mass loss in the inner galaxy within $r_{1/2}$, where Sextans is still DM-dominated \citep[e.g.,][]{wolf_accurate_2010}. The inner mass is only mildly affected with no more than 35 per cent and it's even close to zero in the Light MW model. This modest variance indicates that the dynamical mass of Sextans within $r_{1/2}$ prior to infall does not differ much from its present value, e.g., with 3.6\about5.4$\times10^7$\Msun in our simulations.

By modelling the tidal evolution of Sextans, we can recover the infall virial mass of its DM halo. In the right panel of Fig.~\ref{fig:sim-tidal} we show the virial mass $m_{200}$ of progenitor halo of Sextans recovered by our simulations. The infall virial mass varies a lot across different MW models and orbits, especially in the more massive MW, it can differ with \about$10^9$\Msun when accounting only the orbital pericenter uncertainty. Taking the MW mass into consideration, the infall virial mass is constrained to be no more than $1.5\times10^9$\Msun if the MW mass is $8\times10^{11}$\Msun or less, and no more than $3.1\times10^9$\Msun if the MW mass is $2\times10^{12}$\Msun or less. The properties of DM and stellar component at infall and present time are tabulated in Table~\ref{tab:sim-propt-Sextans} for reference.
\begin{table*}
    \centering
    \caption{
    The simulated properties at infall and present time of Sextans. The first column indicates the orbit model in simulations. The initial parameters to determine the stellar and DM halo model are given in column 2-6. The 7-th column is the bound DM mass that is roughly 1.2 times of virial mass and 8-th column is the circular velocity at $r_{1/2}$ at infall, while 9-th and 10-th are values at the present time. The last column represents the mass loss fraction in DM component, which is defined by $f_{\rm stripping}\equiv1-m_{\rm bound}(t=0)/m_{\rm bound, inf}$.
    }
    \label{tab:sim-propt-Sextans}
    \begin{tabular}{ c |c c c c c c c c| c c | c }
        \hline
        & \multicolumn{7}{c|}{Infall} & & \multicolumn{2}{c|}{Present time} & \\
        \cline{2-8} \cline{10-11}
        Orbit model & $m_{*,0}$ & $a_*$ & $m_{200}$ & $a_{\rm s}$ & ($\alpha, \beta, \gamma$) & $m_{\rm bound,inf}$ & $v_{\rm 1/2,inf}$ &  & $m_{\rm bound}$ & $v_{\rm 1/2}$ & $f_{\rm stripping}$ \\
        & ($\rm 10^5$\Msun) & (kpc) & ($\rm 10^9$\Msun) & (kpc) & & ($\rm 10^9$\Msun) & ($\rm km\,s^{-1}$) &  & ($\rm 10^9$\Msun) & ($\rm km\,s^{-1}$) & (\%)  \\
        \hline
        L81 & 5.94 & 0.522 & 1.46 & 1.171  & (1.79, 2.68, 0.52) & 1.72 & 14.6 &  & 0.86 & 14.1 & 49.9 \\
        L86 & 5.91 & 0.544 & 1.27 & 0.897 & (1.86, 2.65, 0.49) & 1.47 & 14.7 &  & 0.99 & 14.7 & 32.8 \\
        L91 & 5.91 & 0.567 & 1.22 & 0.875 & (1.88, 2.64, 0.48) & 1.41 & 14.7 &  & 1.01 & 14.6 & 28.4 \\
        L81NFW & 5.96 & 0.528 & 0.65 & 1.306 & (1.00, 3.00, 1.00) & 0.75 & 15.2 & & 0.43 & 14.8 & 43.5 \\
        \hline
        M74 & 6.03 & 0.489 & 2.54 & 1.593  & (1.54, 2.79, 0.67) & 2.95 & 16.8 &  & 0.63 & 14.4 & 78.6 \\
        M81 & 6.00 & 0.509 & 2.03 & 1.442  & (1.65, 2.74, 0.61) & 2.37 & 15.9 &  & 0.68 & 14.2 & 71.1 \\
        M88 & 5.99 & 0.565 & 1.66 & 1.166 & (1.74, 2.70, 0.55) & 1.93 & 15.2 &  & 0.64 & 14.0 & 66.7 \\
        M74NFW & 6.16 & 0.513 & 1.14 & 1.605 & (1.00, 3.00, 1.00) & 1.31 & 16.8 & & 0.27 & 15.0 & 79.2 \\
        \hline
        H70 & 6.07 & 0.476 & 3.14 & 1.881 & (1.45, 2.84, 0.73) & 3.64 & 17.8 &  & 0.56 & 14.4 & 84.6 \\
        H78 & 6.06 & 0.503 & 2.35 & 1.600 & (1.58, 2.77, 0.65) & 2.76 & 16.6 &  & 0.58 & 14.1 & 79.0 \\
        H85 & 6.03 & 0.539 & 1.95 & 1.356 & (1.67, 2.73, 0.60) & 2.27 & 15.7 &  & 0.58 & 14.0 & 74.4 \\
        H70NFW & 6.24 & 0.504 & 1.37 & 1.788 & (1.00, 3.00, 1.00) & 1.59 & 17.6 & & 0.25 & 15.2 & 84.4 \\
        \hline
    \end{tabular}
\end{table*}

\subsection{How baryonic effects shape Sextans' properties}
\label{sec:Tidal effect and baryonic effect}
As mentioned in Section~\ref{sec:Dark matter halo}, the baryonic effect on DM profile is incorporated by the $\alpha\beta\gamma$-model from \citetalias{di_cintio_mass-dependent_2014} in our model. To illustrate the separate effect from baryonic and tidal effect on halo density profile, in the upper panels of Fig.~\ref{fig:sim-vcirc} we show the theoretical NFW density profiles assuming virial masses and CMRR concentrations of the L81, M74 and H70 models using red dotted lines (each model is named with a suffix ``(no TE+BE)", such as L81 (no TE+BE) ). The red solid lines represent the baryon modified DM density profiles at infall used by the $\alpha\beta\gamma$-model. As shown, baryonic effects make the inner DM density shallower than the NFW model while remaining the outer part unchanged. Such effects would strongly suppress the inner kinematics while at the same time the large-scale properties almost unchanged. As is indicated by the red dotted and red solid lines in the lower panels of Fig.~\ref{fig:sim-vcirc}, pure baryonic effects reduce the circular velocity at a radius of $r_{1/2}$ by \about3\kms, while the velocities at larger radii are barely affected in our L81, M74 and H70 models.

As mentioned in Section~\ref{sec:Introduction}, however, it is not always true that baryonic feedback will create a shallow density profile. To make a comparison, here we consider three experiments named as L81NFW, M74NFW and H70NFW. In these experiments, the initial DM halo density profile at infall is modeled with NFW profile (see Eq.~\ref{eq:NFW-profile}) and the satellite is placed on the same orbits as in L81, M74 and H70. We follow the same recipes described in Section~\ref{sec:Simulation set-up} to constrain the infall properties of DM and stellar components of Sextans. Therefore, the simulation L81, M74 and H70 account for both of tidal and baryonic effects for Sextans (hereafter TE+BE model) while L81NFW, M74NFW and H70NFW account for tidal effect only (hereafter TE model).

In the upper panels of Fig.~\ref{fig:sim-vcirc} we show the density profiles of both DM and stellar components in these simulations. The solid lines represent the profiles at infall and the dashed lines represent the present time, with red representing the TE+BE model and blue representing the TE model. The lower panels show the circular velocity profiles of the DM component. It can be seen that both the stellar distribution and circular velocity at $r_{1/2}$ are very similar between the TE+BE model and TE model and both of the two models can be reproduced to match the observations at the present time.

However, the halo properties between the two models are largely different. It is obvious in Fig~\ref{fig:sim-vcirc} that the halo mass distribution in the TE model is centrally denser than the TE+BE model and the total mass in the TE model is much smaller than that in the latter, both at the present and infall. The total mass at the present time is \about$4\times10^8$\Msun in simulation L81NFW and \about$2.5\times10^8$\Msun in simulation M74NFW and H70NFW, roughly half of the mass in the TE+BE model.

In Table~\ref{tab:sim-propt-Sextans} we also show the bound DM mass loss fraction for the simulation L81NFW, M74NFW and H70NFW. It is found that the bound DM mass loss fraction between TE+BE and TE model is rather similar since they are on the same orbit. As a result, if the DM halo of Sextans remains the NFW form, the infall virial mass is found to be no more than $6.5\times10^8$\Msun with a MW mass of $8\times10^{11}$\Msun or less, and no more than $1.4\times10^9$\Msun if the MW mass is $2\times10^{12}$\Msun or less. The analysis above shows that the baryonic effect enables Sextans formed in a DM halo with a higher mass.

\section{Discussion}
\label{sec:Discussion}
\subsection{The infall mass of Sextans}
\label{sec:The infall mass of Sextans}
\begin{figure}
    \includegraphics[width=\columnwidth]{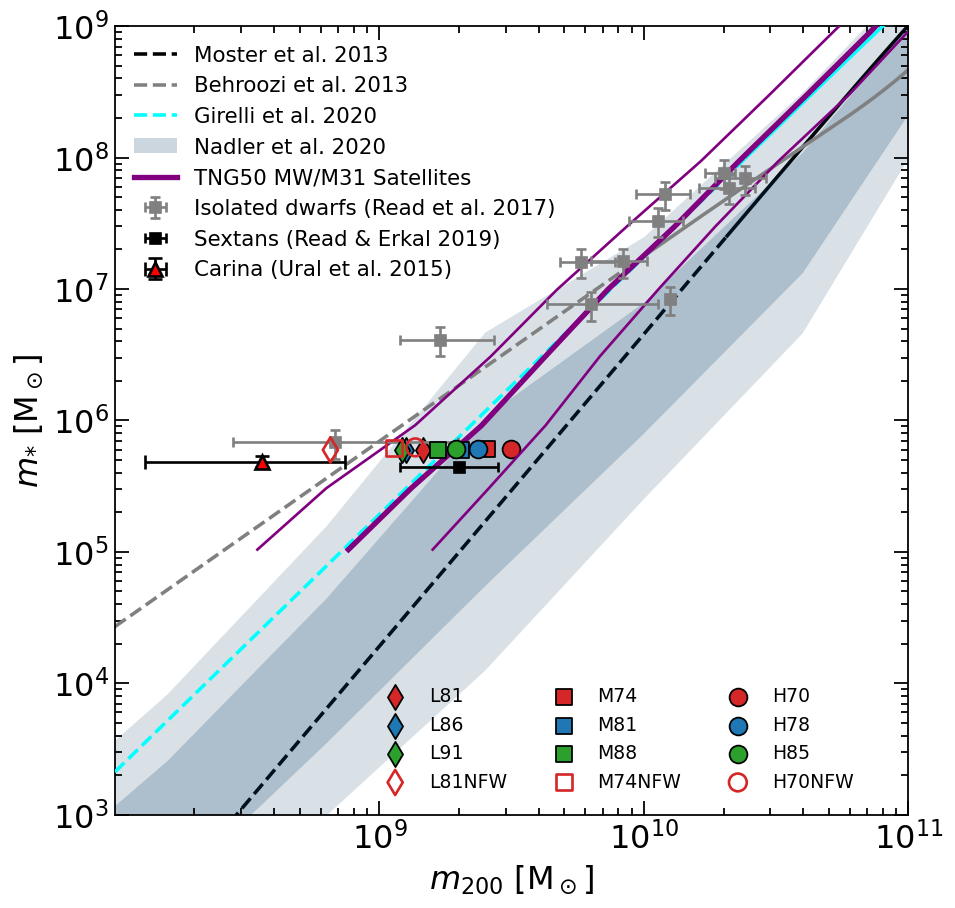}
    \caption{
    The infall virial mass and stellar mass of Sextans compared with SMHM relation at $z=0$ from different empirical models \citep{moster2013,behroozi2013e,girelli2020,nadler2020}. The dashed black, gray and cyan lines are extrapolations of SMHM relations derived using higher-mass galaxies from \citet{moster2013}, \citet{behroozi2013e} and \citet{girelli2020}, respectively. The purple lines depict the SMHM relation of satellite galaxies from TNG50 Milky Way+Andromeda Sample \citep{pillepich_milky_2024} that is calculated at the time when satellites reach their halo mass peak, with the thick line representing the median and thin lines the 16 and 84 percentiles of halo mass at fixed stellar mass down to \about$10^5$\Msun. The gray squres show the SMHM of isolated dwarf galaxies measured by fitting the rotation curves \citep{readStellarMasshaloMass2017c}. We also show individual values with the black square with errorbars representing the halo mass of Sextans based on abundance matching with mean star formation rate \citep{read_abundance_2019} and the red uperr triangle the halo mass of Carina \citep{ural_low_2015}, which is another MW dwarf galaxy with similar stellar mass of Sextans. The dots without errorbars are our simulation results.
    }
    \label{fig:sim-smhm}
\end{figure}
The SMHM relation is a powerful tool to link the stellar mass of galaxies to their DM halo mass \citep[e.g.][]{moster2013,behroozi2013e}. However, the relation at the low-mass end is still not well constrained \citep[e.g.][]{sales_baryonic_2022}. By modelling the tidal evolution of Sextans, we recover its infall virial mass and thus could put constraint on the SMHM relation at the low-mass end.

In Fig.~\ref{fig:sim-smhm} we compare our simulation results with several SMHM relations from the literature. The dashed black, gray and cyan lines are extrapolations of SMHM relations derived by abundance matching using higher-mass galaxies from \citet{moster2013}, \citet{behroozi2013e} and \citet{girelli2020}, respectively. The reason for choosing these models is that the prediction of various galaxy formation models generally lies between these three relations at the low-mass end \citep[e.g.][]{sales_baryonic_2022}. As an example, we also show the SMHM relation of satellite galaxies from TNG50 Milky Way+Andromeda Sample \citep{pillepich_milky_2024} that is calculated at the time when satellites reach their peak of halo mass, which are shown by the purple lines in Fig.~\ref{fig:sim-smhm}. The median SMHM relation in TNG50 follows well the extrapolated relation from \citet{girelli2020}.

As is shown in previous sections, the infall virial mass of Sextans is constrained to be dependent on the MW mass and its DM halo density profile. In the case of a MW mass from $0.8$ to $2\times10^{12}$\Msun, the infall virial mass of Sextans would be in the range of $1.22$\about$3.14\times10^{9}$\Msun if the DM halo density follows the $\alpha\beta\gamma$-model. As shown in Fig.~\ref{fig:sim-smhm}, the halo mass of Sextans follows well with the SMHM relation in TNG50, as well with the extrapolation of \citet{girelli2020}. However, if the DM halo of Sextans remains the NFW form, the infall virial mass is found to be smaller by a factor of \about2, which would tilt the SMHM relation to the extrapolation of \citet{behroozi2013e}. Such a case is not unexpected since the SMHM relation at the low-mass end exhibits large scatter \citep{garrison-kimmelOrganizedChaosScatter2017a,munshiQuantifyingScatterGalaxy2021}. Although the halo mass of Sextans is smaller, it is still consistent with the SMHM of TNG50 within \about$2\sigma$.

For observational constraints, \citet{readStellarMasshaloMass2017c} directly fitted the rotation curves of isolated dwarf irregular galaxies to measure their halo mass. The resulted SMHM relation shows a monotonic and shallower trend with small scatter that follows \citet{behroozi2013e}, as shown by gray squares in Fig.~\ref{fig:sim-smhm}. However, the SMHM relation inferred from MW satellites predicts a steeper trend and large scatter \citep{nadler2020}, as shown by the gray band in Fig.~\ref{fig:sim-smhm}. It is to be seen that whether the relation in \citet{readStellarMasshaloMass2017c} is biased towards the luminous sample. The mass of Sextans would provide helpful insights on the scatter of SMHM relation at the low-mass end. If the virial mass of Sextans is small (when the MW mass is small and the DM halo follows NFW), it would follow well the relation in \citet{readStellarMasshaloMass2017c}, like another MW satellite galaxy Carina (red upper triangle) with similar stellar mass \citep{ural_low_2015}. As a result, the two MW satellites would be typically rare objects in the MW. On the contrary, if the virial mass of Sextans is large (when the MW mass is large and the DM halo follows $\alpha\beta\gamma$-model), it would provide a scatter of \about0.7\dex at fixed stellar mass or \about0.8\dex at fixed halo mass. We note that the abundance matching result using the mean star formation rate of MW satellites supports a larger halo mass for Sextans \citep{read_abundance_2019}, as is shown by the black square with errorbars in Fig.~\ref{fig:sim-smhm}. However, reliable constraints of infall virial mass for more MW satellites would help to draw a firmer conclusion.

\subsection{Uncertainty in observed properties of Sextans}
\label{sec:Uncertainty in observations}
\begin{figure*}
    \includegraphics[width=2\columnwidth]{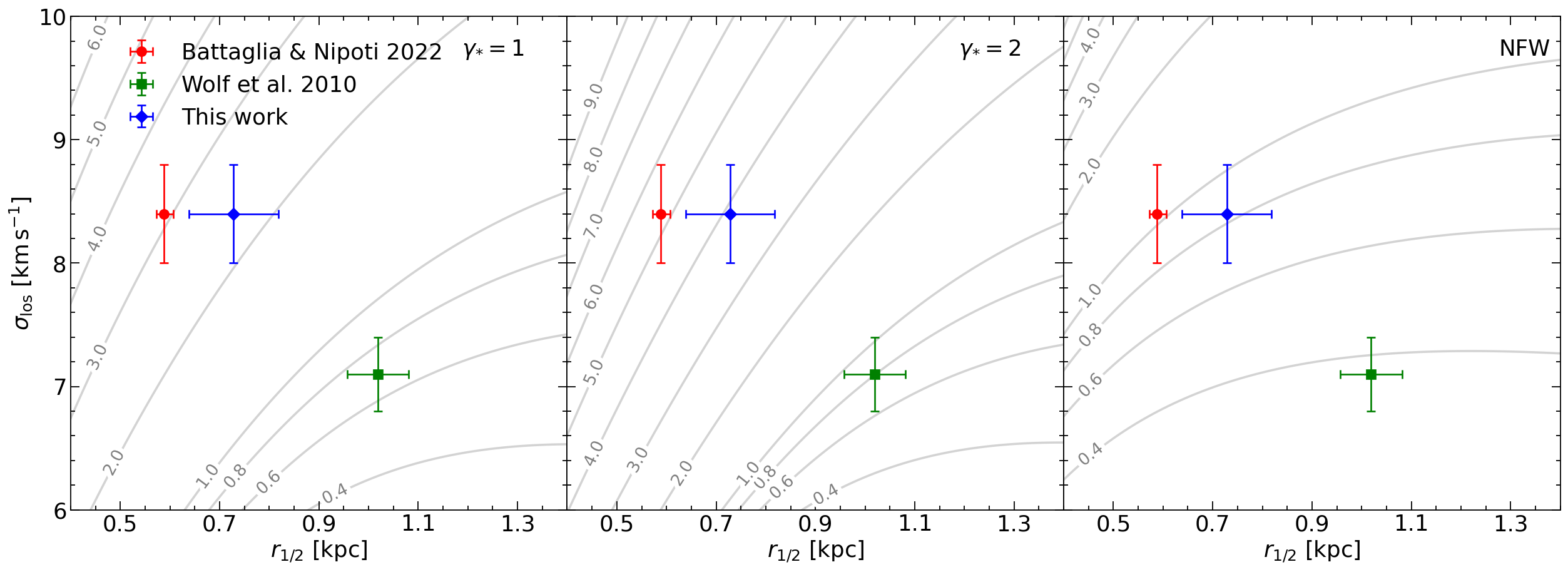}
    \caption{
    The expected infall mass (in unit of $10^9$\Msun) shown by gray lines by varying observed half-light radius, LOS velocity dispersion. The left, middle panels show results by assuming stellar mass-to-light ratio $\gamma_*=1$ and $\gamma_*=2$ with DM halo density following $\alpha\beta\gamma$-model, respectively, while the results for DM halo density following NFW model are shown in the right panel. The data points show three typical $r_{1/2}$ and $\sigma_{\rm los}$ of Sextans from \citet{wolf_accurate_2010} (green squares), and \citet{battaglia_stellar_2022} (red dots), compared with this work (blue diamonds). The error bar represents a $\pm1\sigma$ range.
    }
    \label{fig:sim-unc}
\end{figure*}

\subsubsection{Dynamical mass within $r_{1/2}$}
Over the past two decades, measurements of the DM mass of LG dwarf galaxies have advanced greatly \citep[e.g.][and references therein]{battaglia_stellar_2022}. It has been established that the mass enclosed within the 3D half-light radius can be accurately determined in the fact of flat LOS velocity dispersion profile observed in most classical dSph galaxies of MW \citep{walker_universal_2009,wolf_accurate_2010}. Consequently, the dynamical mass within the half-light radius (hereafter $m_{1/2}$) serves as a fundamental constraint for the DM distribution of these galaxies, as implemented in this work.

However, for the Sextans dSph galaxy, non-negligible uncertainties remain in the structural parameters and kinematics of its stellar component, as mentioned in Section~\ref{sec:The observed properties of Sextans}. In Fig.~\ref{fig:sim-unc}, green squares and red dots represent two typical 3D half-light radii $r_{1/2}$ and global LOS velocity dispersion $\sigma_{\rm los}$ of Sextans from the literatures, which adopt $r_{1/2}$ ranging from $0.588_{-0.016}^{+0.019}$ \citep{battaglia_stellar_2022} to $1.019\pm0.062$\kpc \citep{wolf_accurate_2010} and $\sigma_{\rm los}$ between $7.1\pm0.3$ \citep{wolf_accurate_2010} and $8.4\pm0.4$\kms \citep{battaglia_stellar_2022}. These are compared with our adopted values of $r_{1/2}=0.728\pm0.090$\kpc and $\sigma_{\rm los}=8.4\pm0.4$\kms (blue diamonds).

To understand how these differences affect the infall virial mass determination, we adopt the procedure presented in \citet{kang_warm_2020} to estimate the infall virial mass of Sextans for different $r_{1/2}$ and $\sigma_{\rm los}$ values. In their work, they used the tidal tracks derived by \citet{penarrubiaImpactDarkMatter2010a}
\begin{equation}
    g(x) = \frac{2^{\mu}x^{\eta}}{(1+x)^{\mu}}
\end{equation}
where $x$ is the mass ratio between the present mass and infall mass and $g(x)$ represents either $v_{\rm max}/v_{\rm max,infall}$ or $r_{\rm max}/r_{\rm max,infall}$, and they interpolate the parameters $\mu$ and $\eta$ for different inner density slopes $\gamma$ to model the tidal evolution of MW satellites with various DM halo density profiles. We note that in this procedure, the inner density slope $\gamma$ is constant during the tidal evolution while the outermost region scales as $r^{-5}$, which is a good approximation according to \citet{penarrubiaImpactDarkMatter2010a}. Once we know the mass distribution for a given DM halo, the stellar kinematics $\sigma_{\rm los}$ is simply linked to the circular velocity at $r_{1/2}$ by $v_{1/2}=\sqrt{3}\sigma_{\rm los}$ \citep{wolf_accurate_2010}. 

Drawing from our simulation results in Table~\ref{tab:sim-propt-Sextans}, the tidal mass loss fraction for Sextans varies from 28.4 to 84.6 per cent depending on the MW mass and its orbit, with weak dependence on the halo density profile for given orbit. Therefore, in our deduction of the infall mass, we use a typical tidal mass loss fraction of 71.1 per cent which is the value in simulation M81. The infall virial mass of Sextans is estimated by using both of the $\alpha\beta\gamma$-model and the NFW model. Since the stellar mass is not obviously affected by tides, we assume a fixed stellar mass of $5.9\times10^5$\Msun in calculation. The infall time is simply set to be $z_{\rm inf}=1$ and we have checked that varying $z_{\rm inf}$ does not affect the results very much.

The resulting infall virial masses $m_{200}$ are shown as gray lines in Fig.~\ref{fig:sim-unc}. Basically, a smaller half-light radius and larger velocity dispersion would bring the infall mass larger. The left panel of Fig.~\ref{fig:sim-unc} shows the results with the $\alpha\beta\gamma$-model assuming a stellar mass-to-light ratio $\gamma_*=1$, which is the same case adopted by our simulation. The blue diamond locates at an infall mass of $2.3\times10^9$\Msun which is slightly larger yet still consistent with the results of simulation M81. The smaller half-light radius adopted by \citet{battaglia_stellar_2022} yields a larger infall virial mass of $3.1\times10^9$\Msun. However, the extended $r_{1/2}$ and lower $\sigma_{\rm los}$ values in \citet{wolf_accurate_2010} produce a significantly smaller infall virial mass estimate of $6.9\times10^8$\Msun.

For the NFW model, the results are shown in the right panel of Fig.~\ref{fig:sim-unc}. The infall mass is remarkably lower than those following $\alpha\beta\gamma$-model by a factor of \about2-3. Again the results are consistent with our simulation in the TE+BE model and TE model presented in Section~\ref{sec:Tidal effect and baryonic effect}.

\subsubsection{Stellar mass}
Another important factor influencing the determination of virial mass of Sextans is its stellar mass since baryonic effects are dependent on the stellar-to-halo mass ratio \citep[e.g.,][]{di_cintio_dependence_2014,di_cintio_mass-dependent_2014}. In our simulations, we adopt a fixed stellar mass of $5.9\times10^5$\Msun, which is relatively conservative in reshaping the inner DM density slope with a minimum cored density scaled as $r^{-0.5}$ in the inner region. It is natural according to the \citetalias{di_cintio_mass-dependent_2014} model that larger stellar masses will produce shallower inner density profiles.

Therefore, the middle panel of Fig.~\ref{fig:sim-unc} shows how varying the stellar mass of Sextans affects the results. We assume a larger stellar mass-to-light ratio with $\gamma_*=2$, resulting a stellar mass of $1.18\times10^6$\Msun, which is close to the upper limit of stellar mass estimated by \citet{woo_scaling_2008} and \citet{karlsson_chemical_2012}. Compared with the gray lines in the left panel, the resulting infall virial mass in the middle panel shows a boost by a factor of \about1.2-1.7.

However, if the DM halo density is not affected by baryonic effects and remains the NFW form, varying the stellar mass would not change the virial mass determination in our simulation. We thus do not show the results again.

\subsection{The DM halo density profile of Sextans}
\label{sec:The DM halo density profile of Sextans}
\begin{figure}
    \includegraphics[width=\columnwidth]{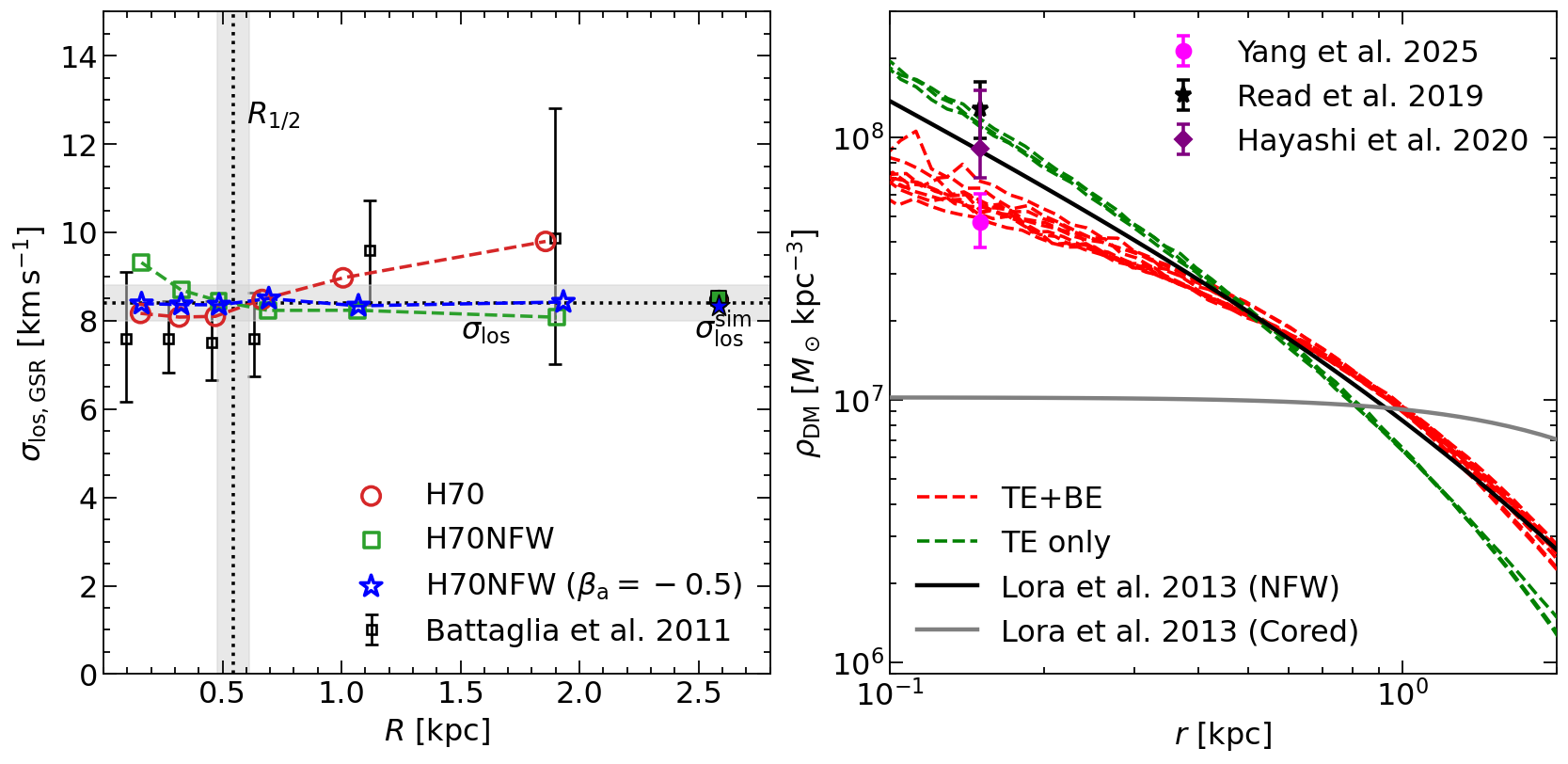}
    \caption{
    A comparison between simulations with the $\alpha\beta\gamma$-model (red lines) and the NFW model (green lines).
    The left panel shows the stellar LOS velocity dispersion profiles at present time in simulation H70 (red circles), H70NFW (green squares) and an H70NFW variant with a constant anisotropic parameter $\beta_{\rm a}=-0.5$ (blue stars). The black squares represent the observation. The right panel shows the DM halo density profiles from 0.1 to 2\kpc at present time in all of our simulations. The red dashed lines represent results of 9 simulations for the TE+BE model. Green lines represent results of 3 simulations for the TE model. The black and gray solid lines in this panel show the density profile of the NFW and cored DM halo modelled in \citet{lora_sextans_2013}. The black star, purple diamond and magenta dot show the DM density at 150\pc of Sextans derived by \citet{read_dark_2019}, \citet{hayashi_diversity_2020} and \citet{yang_dark_2025}, respectively. The data are from \citet{yang_dark_2025}.
    }
    \label{fig:sim-density}
\end{figure}
In this section, we simply compare the difference of the properties of Sextans between simulations using different DM halo density profiles.

As is shown in previous sections, both of the $\alpha\beta\gamma$-model and the NFW model are able to reproduce the observational constraints in our simulations. However, the shape of the LOS velocity dispersion profile between the two models are different. As an example, in the left panel of Fig.~\ref{fig:sim-density} we compare the LOS velocity dispersion profile between the H70 simulation and the H70NFW simulation. The global LOS velocity dispersions within $1.6$\degree are identical between the two simulations (shown by the filled red circle and green squares at $R\sim2.5$\kpc), which are consistent with the observation. However, the cuspy NFW model produces a declining LOS velocity dispersion profile which is in contradiction with the observed increasing profile by \citetalias{battaglia_study_2011}. It is also found from the plot that our $\alpha\beta\gamma$-model better reproduces the LOS velocity profile of Sextans. \citetalias{battaglia_study_2011} suggested that the LOS velocity dispersion profile is affected by the tangential anisotropy parameter ($\beta_{\rm a}$). Here we show a case in which $\beta_{\rm a}=-0.5$ is adopted for the H70NFW model, that is shown by the blue stars. It is found that introducing a tangential velocity anisotropy will flatten the LOS velocity profile, leading to better agreement with the data.

A common indicator of central DM density is $\rho(150\rm\,pc)$, which is the DM density at 150\pc \citep[e.g.,][]{read_dark_2019,hayashi_diversity_2020,yang_dark_2025}. In the right panel of Fig.~\ref{fig:sim-density}, we show the DM density profiles within 2\kpc at present time for all of our simulations, including 9 simulations in the TE+BE (red dashed lines) and 3 simulations in the TE model (green dashed lines). As expected, the TE model gives larger $\rho(150\rm\,pc)$ compared to the TE+BE model. Based on our simulations, the baryon-induced core and a cuspy NFW profile can be distinguished with a factor of 2 times variation between these models. However, present constraints for $\rho(150\rm\,pc)$ using different observed samples are still too uncertain \citep[e.g.,][]{yang_dark_2025}. Here we compare the $\rho(150\rm\,pc)$ of Sextans derived by different work which are shown by the points with errorbars in the right panel of Fig.~\ref{fig:sim-density}. It can be found that both cuspy and cored profiles are consistent with these work.

Globular clusters may serve as better probes of central density profiles \citep{kleyna_dynamical_2003,lora_sextans_2013,meadows_cusp_2020,modak_distinguishing_2023}. Several lines of evidence suggest star cluster remnants exist in Sextans, including kinematically cold substructures (\citealp{kleyna_photometrically_2004};\citetalias{battaglia_study_2011}), chemical coherence of stars \citep{karlsson_chemical_2012}, and number density excess \citep{kim_possible_2019} in the central region. \citet{lora_sextans_2013} demonstrated that cuspy halos would disrupt globular clusters within \about$5$\Gyr while cored halos preserve them. They adopted a DM halo mass of $2.6\times10^9$\Msun and concentration of $10$ following \citetalias{battaglia_study_2011} for the cuspy model, which is shown by the black solid line in the right panel of Fig.~\ref{fig:sim-density}. It can be seen that the inner density of this model is denser than that in our TE+BE models while shallower than our TE models. Therefore, the survival of globular clusters in Sextans would challenge the cuspy density profiles that can disrupt possible stellar clumps. However, all of our models are denser than the cored model in \citet{lora_sextans_2013}. It remains to be seen that whether such models would coincide with all of the observational evidence in Sextans. The structural parameters derived by our simulations provide a valuable baseline for future investigations of this system.

\section{Summary}
\label{sec:Summary}
In this work, we investigate the tidal history for the MW satellite galaxy Sextans. By exploring possible orbits of Sextans in different MW potentials using updated PM measurements from \textit{Gaia} EDR3 \citep{li_gaia_2021}, we conduct tailored $N$-body simulations to recover the infall properties of its DM halo. All of our simulations reproduce the observed stellar mass, half-light radius \citep{wolf_accurate_2010,roderick_structural_2016,okamoto_population_2017,cicuendez_tracing_2018,tokiwa_study_2023} and LOS velocity dispersion \citepalias{battaglia_study_2011} of Sextans at present time. We consider two different DM halo density profiles for Sextans at infall, one of which is the $\alpha\beta\gamma$-model provided in \citetalias{di_cintio_mass-dependent_2014} that incorporates the baryonic effects and the other is the NFW model. 

These simulations enable, for the first time, a detailed examination of how baryonic and tidal effects shape Sextans' properties. The tidal effects on Sextans are rather simple with its stars not or only mildly affected by galactic tides and the stellar kinematics provide robust constraints on the dynamical mass within $r_{1/2}$. The tidal stripping on the DM halo of Sextans is rather uncertain mainly due to the uncertainty of the MW mass, ranging from 28 per cent mass loss in our Light MW model to 85 per cent in the Heavy MW model.

We find that baryonic effects produce a limited baryonic core in the inner DM density of Sextans with a minimum logarithmic slope of about $-0.5$ in the central region. Even though, baryonic effects are still important to suppress the inner kinematics of Sextans by reducing the circular velocity at $r_{1/2}$ by \about$3$\kms. Such effects allow Sextans to form in a higher mass DM halo from $1.22$ to $3.14\times10^9$\Msun for a MW mass from 0.8 to 2$\times10^{12}$\Msun. By contrast, if the initial DM density of Sextans is cuspy like NFW, the infall mass recovered based on the present observational constraints would be smaller by a factor of 2.

We also discuss the implications of uncertainty of observed properties of Sextans on its infall mass. Basically a smaller half-light radius, larger velocity dispersion or higher stellar mass would result in a larger infall mass. Nevertheless, the possible infall masses of Sextans recovered by our simulations are consistent with the SMHM relation in TNG50 and abundance matching results. We find some peculiar cases for the NFW DM density profile where the infall mass of Sextans can be smaller than $10^9$\Msun if the MW mass is small, or the half-light radius of Sextans is large and the velocity dispersion is small \citep[e.g.][]{wolf_accurate_2010}. Like the case of Carina \citep{ural_low_2015}, such a low halo mass would make them typical rare objects in the MW and indicate the stochasticity of galaxy formation at the dwarf galaxy scale. We suggest reliable constraints of infall halo masses for more MW satellites are helpful to constrain the scatter of SMHM relation at the low-mass end.

The inner DM halo density profile of Sextans is thus important to understand not only the nature of DM, but also the baryonic physics in dwarf galaxies since the SMHM relation at the low-mass end varies in different galaxy formation models \citep[e.g.,][]{sales_baryonic_2022}. However, the inner density of Sextans is still not well constrained by present observations. An increasing LOS velocity dispersion profile is consistent with either a cored density profile or a cuspy density profile with tangential anisotropy \citepalias{battaglia_study_2011}. We note that varying the anisotropic parameter does not influence the infall virial mass determination in our simulations. The possible existence of globular clusters in Sextans may provide better probes of its central density profile \citep{lora_sextans_2013,meadows_cusp_2020,modak_distinguishing_2023}. The structural parameters derived by our simulations could inform future studies of Sextans' central structure.

\section*{Acknowledgements}
We thank Yaoxin Chen for the useful discussions about the halo mass evolution in Eq.~\ref{eq:mwacc-mass}, and the anonymous referee for the valuable comments. This work was supported by the National Natural Science Foundation of China (NSFC, No. 12533007, 12595314, 12547104, 12347103, 12273027, 12373004, W2432003, 12573024), the National Key Research and Development Program of China (No.2022YFA1602903, No.2023YFB3002502, No.2023YFA1608100), the science research grants from the China Manned Space project with No. CMS-CSST-2025-A10, CMS-CSST-2025-A20, the Fundamental Research Funds for the Central Universities of China (226-2022-00216) and the start-up funding of Zhejiang University. The simulations and data analysis were performed on the SilkRiver high-performance computing platform at Zhejiang University.

\section*{Data Availability}
The data underlying this article will be shared on reasonable request to the corresponding author.


\bibliographystyle{mnras}
\bibliography{refs} 




\appendix


\bsp	
\label{lastpage}
\end{document}